\documentclass[%
preprint,
amsmath,amssymb,
aip,
]{revtex4-1}

\usepackage{color}
\usepackage{graphicx}
\usepackage{dcolumn}
\usepackage{bm}
\usepackage[section]{placeins}

\newcommand{\RE}{{\text{Re}}}
\newcommand{\FRO}{{\text{F}_\nu}}

\begin{document}
	

\title{Modeling compressed turbulent plasma with rapid viscosity variations}
\author{S\'ebastien Th\'evenin}%
\affiliation{%
	CEA, DAM, DIF, F-91297 Arpajon, France 
}%
\author{Nicolas Valade}%
\affiliation{%
	CEA, DAM, DIF, F-91297 Arpajon, France 
}%
\author{Beno\^it-Joseph Gr\'ea}%
\affiliation{%
	CEA, DAM, DIF, F-91297 Arpajon, France }
\author{Gilles Kluth}%
\affiliation{%
	CEA, DAM, DIF, F-91297 Arpajon, France 
}%
\author{Olivier Soulard}%
\affiliation{%
	CEA, DAM, DIF, F-91297 Arpajon, France 
}%
\date{\today}
\begin{abstract}
	We propose two-equations models in order to capture the dynamics of a turbulent plasma undergoing compression and experiencing large viscosity variations.   
	The models account for possible relaminarization phases and rapid viscosity changes through closures dependent on the turbulent Reynolds and on the viscosity Froude numbers. These closures are determined from a data-driven approach using eddy-damped quasi normal markovian simulations. The best model is able to mimic the various self-similar regimes identified in \citet{Viciconte2018} and to recover the rapid transition limits identified by \citet{Coleman1991}.
\end{abstract}

\pacs{Valid PACS appear here}
\maketitle

\section{INTRODUCTION}

Turbulent mixing is a key element for determining the yield of an inertial confinement fusion (ICF) capsule. High $Z$ elements from the ablator (mostly CH) and mixing with the deuterium-tritium (DT) fuel may deteriorate the rate of fusion reactions by cooling the hot spot. This is principally induced by an enhancement of conductivity and X-ray emissions. Turbulent mixing may be generated by different mechanisms such as the classical Rayleigh-Taylor and Richtmyer-Meshkov hydrodynamic instabilities initiated at the fuel/ablator interface \cite{Clark2015,Remington2018}. Developing efficient turbulence models is therefore of paramount importance for the design of ICF capsules \cite{Campos2019a}. 

Turbulence encountered during the compression of an ICF plasma is specific in many aspects. When the plasma gets close to the kinetic regime, at the end of the compression for instance, plasma models suggest that the transport coefficients experience a tremendous growth \cite{Braginskii1965,Clerouin2020}. As a consequence, the dissipation of turbulent kinetic energy becomes not only driven by the cascading process expressed by a turbulent eddy viscosity but also directly by the plasma viscosity $\nu$. Eventually, a sudden relaminarization of the flow may occur dissipating entirely the turbulent kinetic energy \cite{Davidovits2016}. Whether or not this effect is dominant is often difficult to assess. However in many situations, both the turbulent and the plasma diffusions due to transport coefficients drive the mixing in ICF capsules \cite{Weber2014,Haines2016,Zylstra2018,
Viciconte2019,Campos2019b,Mackay2020,Haines2020}.

Additionally, the very strong unsteadiness encountered in ICF turbulent plasma is by itself a critical element to predict the dissipation. \citet{Lumley1992} has summarized this well-known issue: 
'What part of modeling is in serious need of work? Foremost, I would say, is the mechanism that sets the level of dissipation in a turbulent flow, particularly in changing circumstances'.
Most turbulence models rely indeed on the link between the dissipation $\varepsilon$ and the energy containing eddies characterized by their velocity $\mathcal U$ and scale $\mathcal L$.  This gives the fundamental relationship first proposed by \citet{Taylor1935}
\begin{equation}
	\varepsilon = C_\varepsilon \frac{\mathcal U ^3}{\mathcal L}, \label{eq:C}
\end{equation}
where $C_\varepsilon$ is the renormalized dissipation rate usually considered as constant. Due to Eq.~(\ref{eq:C}), the variable $\varepsilon$ implicitly acquires a double meaning, as it expresses not only the dissipation rate but also the energy transfer from the large to the small scales due to the cascade. This approach is therefore very successful in many practical situations where a spectral equilibrium is achieved and the classical Kolmogorov phenomenology applies. However, when the energy transfer does not equal the dissipation as in very unsteady flows, Eq.~(\ref{eq:C}) is no longer justified (see also  \cite{Tennekes1972, Vassilicos2015}).  

A first step toward predicting turbulent plasmas under compression has been successfully proposed by \cite{Davidovits2017}. Their model is based on a dependence of the coefficient $C_\varepsilon$ with the Reynolds number adaptated from \cite{Loshe1994}. Noticeably, this is a one equation approach, rendered possible when the integral scale $\mathcal L$ remains fixed in the frame attached to the compression. In many practical configurations however it is not the case and a model with at least two equations is requested.

Therefore, the objective of this work is to construct a model for turbulent plasma under compression which can recover both the self-similar solutions detailed \cite{Viciconte2018} and the rapid phases identified in \cite{Coleman1991}. In doing so, it is important to limit the complexity of the model, particularly to allow the extension of existing models to plasma under compression with few modifications. This is why we will restrict our analysis to two equations models  classically used by engineers. 

This paper is organized as follows. First, we detail the eddy-damped quasi normal markovian (EDQNM) simulations and the various rapid or relaminarization phases existing in a turbulent plasma under compression. Then we provide various strategies to model such a flow from classical models to fully data-driven closures.  

\section{The dynamics of a turbulent plasma under compression}

In this section, we detail the basic equations describing the dynamics of a turbulent velocity field in a plasma under compression. Assuming statistical homogeneity and isotropy, an EDQNM model for second order correlation spectra can be derived, allowing the exploration of the different regimes encountered during the compression. In particular, we emphasize the self-similar and rapid phases driven by the sudden growth of the viscosity coefficient, in either a laminar or turbulent context.

\subsection{Basic equations}

In order to develop a turbulence model accounting for viscosity variations, it is convenient to work with well-characterized numerical solutions which can be provided by the idealized framework of compressed homogeneous isotropic turbulence. 
We thus consider a turbulent plasma initially occupying a given volume and compressed by a radial velocity on the form ${\bf U}=-\mathcal S_0 \bf {r}$. Here $1/\mathcal S_0$ is the characteristic time of the compression and $\bf r$ the radial unity vector. Classically, the fluctuating turbulent velocity field $\bf u$ of the plasma can be expressed in the frame attached to the compression.  This simply reduces to the Navier-Stokes equations as (see details in \cite{Nishitani1985, Cambon1992,Cambon1993,Davidovits2016, Viciconte2018})
\begin{subequations}
	\begin{align}
	&\partial u_i+\partial_j (u_i u_j)=-\partial_i p+ \nu \partial _{jj} u_i \label{eq:basics1},
	\\
	&\partial_j u_j=0 \label{eq:basics2}.
	\end{align}
\end{subequations}

Although the plasma is obviously fully compressible, Eq.~(\ref{eq:basics2}) expresses that the turbulent fluctuations stay incompressible which seems a good first approximation in the context of ICF plasmas (see \cite{Viciconte2019}). Put in another way, the internal energy is assumed much larger than the turbulent kinetic energy. In Eq.~(\ref{eq:basics1}), the kinematic viscosity varies in the frame attached to the compression as $\nu=\nu_0 ( \mathcal S_0 t+1)^\theta$, introducing a reference initial viscosity $\nu_0$. The growth exponent is supposed positive, $\theta>0$, as the viscosity keeps increasing in the ICF context. It is also maximum in the kinetic regime corresponding to the Braginskii law \cite{Braginskii1965} and for an isentropic compression with $\theta=2$. In addition, the viscosity expression indicates the dependence of the transport coefficient to the density and the temperature evolving during the compression. Therefore the compression parameters are hidden in the evolution of the viscosity $\nu$ together with the change of variable from the laboratory to the compression frame. Studying the dynamics of a compressed plasma is here equivalent to studying the decay of unforced turbulence with a time varying viscosity coefficient.

\subsection{EDQNM simulations}

Homogeneous isotropic turbulence with time growing viscosity exhibits large energy and length variations. Exploring it with direct numerical simulations (DNS) can be difficult as the simulations, in addition to being costly, may suffer from an under-resolution of small scales and a confinement of larger ones due to the finite size of the mesh. It is therefore convenient to use an EDQNM model which, by providing the dynamics of energy spectra $E(k)$ on a very large panel of wavenumbers $k$, is sufficient to recover one point quantities classically used in RANS models. 
The EDQNM model in this study follows the original method proposed by \cite{Orszag1970} but is adapted to turbulence in compression as detailed by \cite{Viciconte2018}. 
This model has proven its ability to reproduce a large panel of DNS data at low cost. For this study, 143 EDQNM simulations have been used.  

The EDQNM simulations are initialized with a von Karman energy spectrum of the form $E(k)\sim k^s \exp \left[-s/2 (k/k_\text{peak})^2\right]$, with $s$ the slope of the infrared spectrum and $k_\text{peak}$ the wavenumber corresponding to the maximum of the energy spectrum. In this study, we mainly focus on $s=2$ corresponding to Saffman spectra commonly encountered in experiments. 
In a preliminary stage, corresponding to $t < 0$, we keep the viscosity constant and let the simulations evolve for several eddy turnover times in order to obtain fully developed spectra. We check that at $t=0$ (this indeed defines the initial condition), the simulations are nearly in a classical self-similar decay regime with the kinetic energy evolving as $K=\int_0 ^{+\infty} E(k) dk \sim t^{-n_T}$. Here the decay exponent is expected to take the value $n_T=2 (s+1)/(s+3)=6/5$ for sufficiently turbulent initial conditions \cite{Lesieur2000}. Then for $t>0$, the viscosity coefficient can grow freely with the exponent $\theta$.

\begin{figure}
	\begin{center}
		\includegraphics[width=0.6\textwidth]{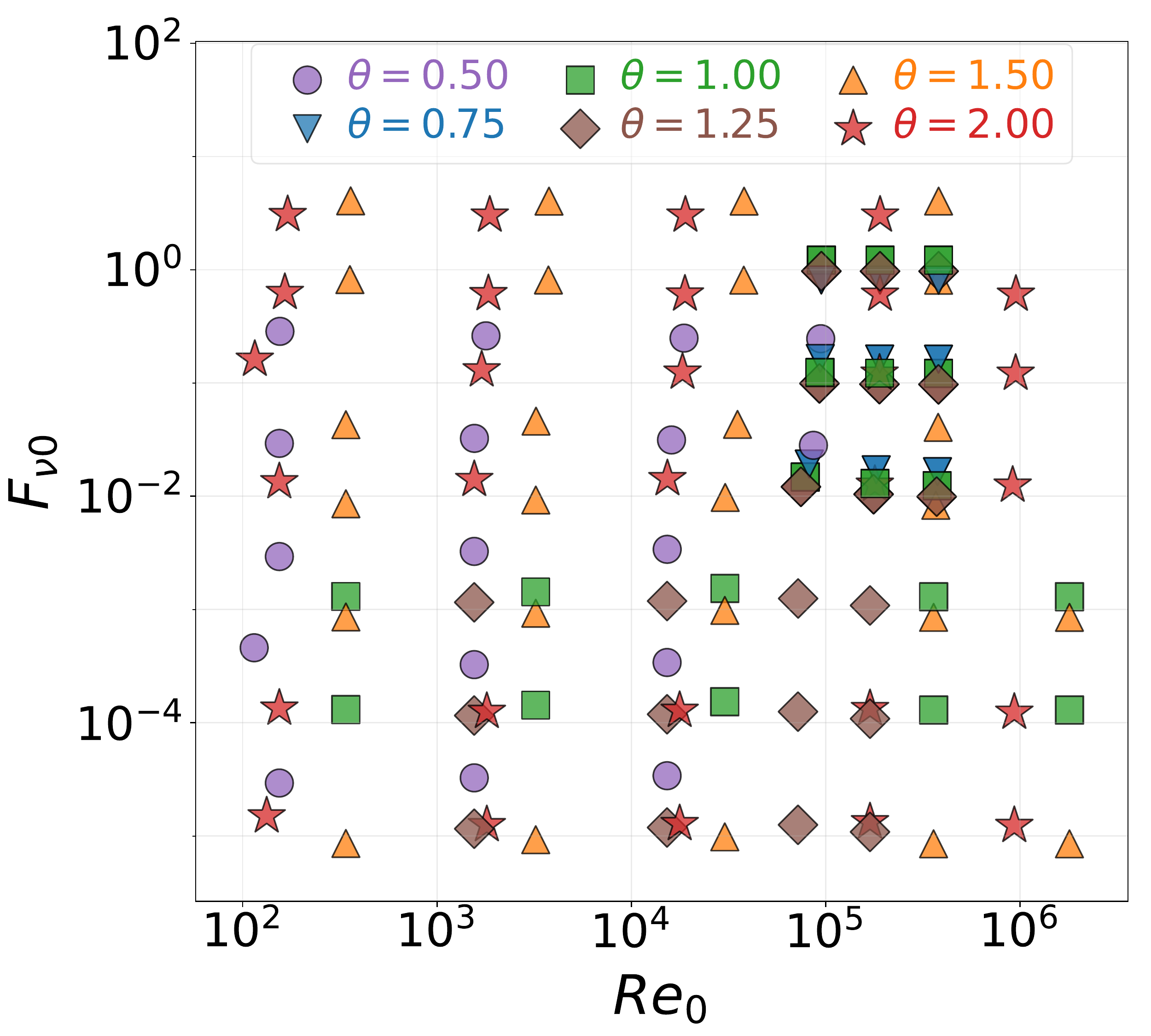}
		\caption{Distribution of the initial conditions ($t=0$) for the EDQNM simulations in term of Reynolds $Re$ and Froude $F_\nu$ numbers. The symbols indicate to the various viscosity growth exponent values $\theta$.
		These conditions correspond to fully developed turbulent spectra. \label{fig:init}}
	\end{center}
\end{figure}

The turbulent Reynolds $\RE$ and Froude $\FRO$ numbers fully characterize the EDQNM simulations. They are given by
\begin{align}
	\RE=\frac{K^2}{\varepsilon \nu} \ \text{and} \
	\FRO=\frac{\varepsilon}{K} \frac{\nu}{\dot \nu}, \label{eq:refr}
\end{align}
where the dissipation $\varepsilon$ is obtained from the energy spectrum as $\varepsilon=2 \nu \int_0 ^{+ \infty} k^2 E(k,t) dk $.  In  Eq.~(\ref{eq:refr}), the turbulent Reynolds number expresses the size of the inertial range of the spectrum. In addition, the Froude number indicates the intensity of the viscosity growth as 
the ratio between the turbulent frequency $\varepsilon/K$ and the viscosity growth rate $\dot \nu / \nu$. Therefore, a small Froude number indicates that the viscosity growth is strong and can have an impact on the 
turbulence.
We present in the Figure~\ref{fig:init} the various EDQNM simulations in terms of initial Reynolds and Froude numbers. 
These simulations exhibit a large panel of Froude and Reynolds numbers corresponding to a turbulent regime.

\begin{figure}
	\begin{center}
		\includegraphics[width=0.8\textwidth]{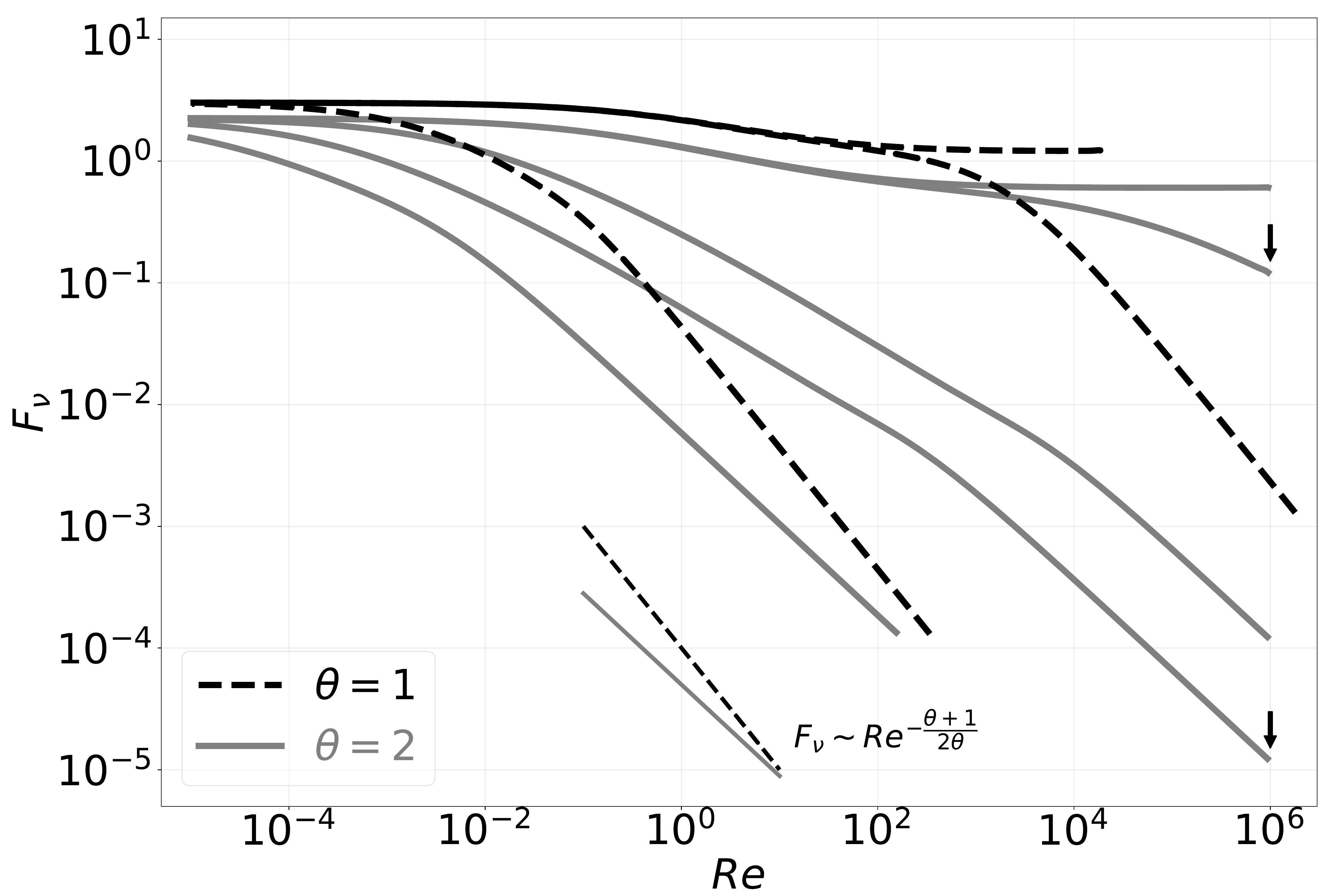}
		\caption{Trajectories for selected EDQNM simulations in the Reynolds $Re$ and Froude $F_\nu$ plane. The line types indicate different $\theta$ values. As the time grows, the Reynolds number decreases and the Froude number converges to a constant when the flow enters the viscous self-similar regime. We also show the slopes corresponding to the asymptotic trajectories, {\it i.e.} $\FRO \sim \RE ^{-\frac{\theta+1}{2 \theta}}$, in the rapid viscous phases. The arrows indicate the two EDQNM simulations analyzed in Figure~\ref{fig:ss}.
		\label{fig:evol}}
	\end{center}
\end{figure}

We also show in Figure~\ref{fig:evol} typical evolution of the Reynolds and the Froude numbers in selected EDQNM simulations. 
As the time grows, the Reynolds number decreases  due to the viscous dissipation which is not balanced by any production terms. Noticeably, more than $10$ orders of magnitude can be captured by the EDQNM simulations, which would have been impossible to obtain from a DNS. Although the initial Froude numbers take very different values at large Reynolds numbers, the Froude numbers become constant at low Reynolds numbers. This is the imprint of the viscous self-similar regime as will be discussed in the next section. It should be stressed that an asymptotic value for $F_\nu$ in the viscous self-similar regime seems difficult to predict. One can also notice that the trajectories for different $\theta$ values intersect, particularly in the small Froude number regions. This aspect indicates a dependence of the results to the viscosity history, which is particularly daring during rapid viscosity phases as detailed too in the following section.      

\subsection{Self-similar regimes and rapid viscous phases \label{sec:selfsimilar}}

The decay of a turbulent flow with time-varying viscosity is often marked by a turbulent and a viscous phase, both eventually evolving self-similarly. For the turbulent regime, the self-similar decay exponent is only determined  by the large scale properties of the flow and takes the value $n_T=6/5$ for Saffman spectra as already mentioned \cite{Lesieur2000}. 
In the viscous self-similar regime, the kinetic energy is expected to decay as $K \sim t^{-n_v}$. The decay exponent has been derived in \cite{Viciconte2018} for $\theta=2$ (plasma under isentropic compression in the kinetic regime) leading to $n_v=3(s+1)/2$. It is possible to generalize this formula using the same arguments, namely the permanence of large eddies and the fact that the integral scale evolve as $\sim \nu(t) K /\varepsilon$, giving $n_v=(1+\theta)(s+1)/2$. As expected, the decay is more pronounced in the viscous regime as $n_v> n_T$.

The self-similar viscous decay exponent, $n_v$, also determines whether or not the plasma under compression experiences a rapid viscous dissipation phase as described in \citep{Davidovits2016}. Accounting for the transformation between the frame attached to the compression and the laboratory frame, the relaminarization condition writes $n_v \ge 2$. This implies a growth rate exponent of the viscosity of $\theta \ge (3-s)/(s+1)$. The result expresses an important sensitivity to the large scale distribution of energy, leading for instance to $\theta \ge 1/3$ for $s=2$. Noticeably, this condition corresponds to a dynamic viscosity growth with the temperature $T$ of $\sim T^{5/3}$ for an isentropic compression. This differs from the condition established in \citep{Davidovits2016b} of $\sim T$  as this latter analysis assumes the confinement of turbulence at large scales.

\begin{figure}
	\begin{center}
		\includegraphics[width=0.7\textwidth]{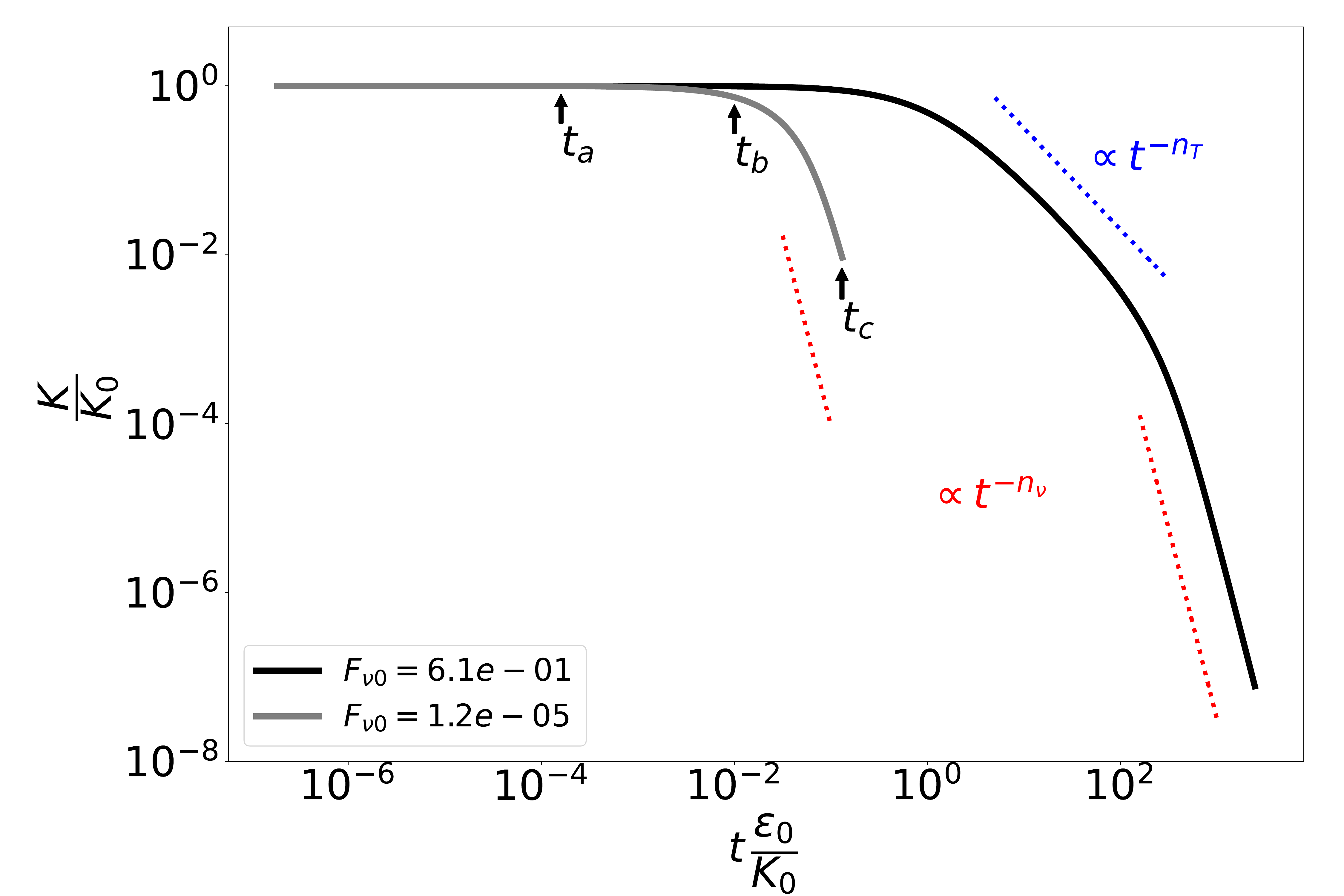}
		\includegraphics[width=0.7\textwidth]{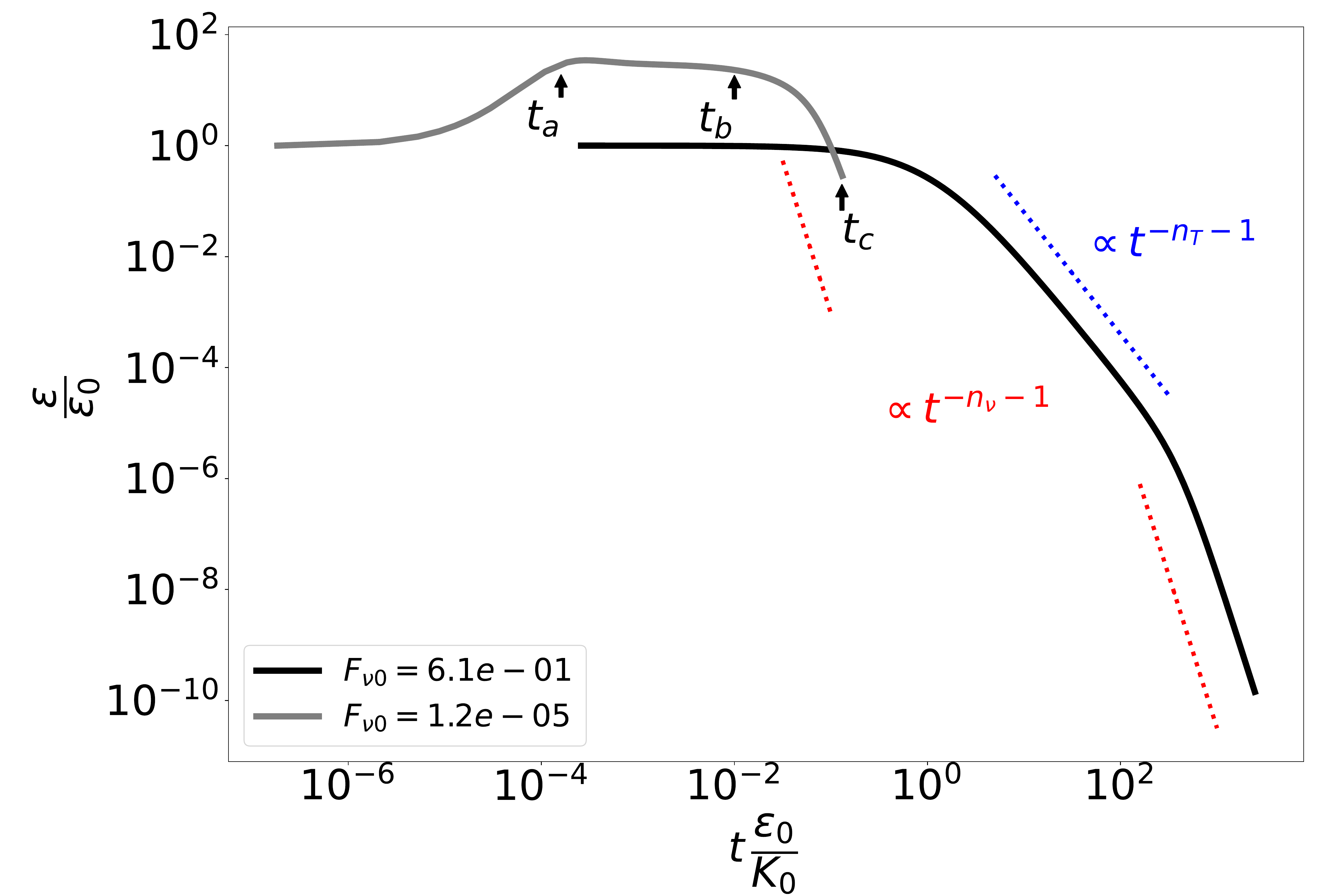}
		\caption{ Time evolution of kinetic energy (Top) and 
			dissipation (Bottom) for two EDQNM simulations with $\theta=2$, initial Reynolds number $\RE _0 \sim 10^6$ and initial Froude numbers $\FRO _0$ indicated in insert. The blue and red slopes associated with exponents $n_T$ and $n_v$ correspond to the theoretical turbulent and viscous self-similar regimes respectively. The times $t_{a-c}$ correspond to the spectra in Figure~\ref{fig:spec}.\label{fig:ss}}
	\end{center}
\end{figure}

We can now verify that these scaling arguments are recovered in the EDQNM simulations. In Figure~\ref{fig:ss}, we present the time evolution (renormalized by the initial turbulent frequency) of the kinetic energy and the dissipation in two simulations with $\theta=2$ at initially $\RE _0 \sim 10^6$ and $\FRO _0 \sim 0.6$ or $1.2 \times 10^{-5}$. The trajectories of these two representative simulations in the $(\RE,\FRO)$ plane are also visualized in Figure~\ref{fig:evol}.
Both simulations converge to a self-similar viscous decay as shown by the scaling of the turbulent energy, $K \sim \ t^{-n_v}$, and the dissipation, $\varepsilon=-\dot K\sim t^{-n_v-1}$. This phase corresponds to the plateau in term of Froude number in Figure~\ref{fig:evol} at low Reynolds number. Interestingly, a sharp difference between the two simulations occurs in the high Reynolds number region. The simulation with initially $\FRO _0 \sim 0.6$ exhibits a self-similar turbulent regime which is also characterized by a plateau of Froude number. By contrast, the low Froude number simulation jumps directly into the viscous phase with a quick decay of kinetic energy and a transient growth of the dissipation. This phenomenon, hereafter referred as the rapid viscosity phase, has a strong importance in the dynamics of turbulent quantities.  

In order to get insight of this rapid viscosity phase, we first recall the equations for the kinetic energy and the dissipation (see for instance \cite{Lesieur1990}):
\begin{subequations}
	\begin{align}
	&\frac{d K}{dt}=-\varepsilon, \label{eq:K} \\
	&\frac{d \varepsilon}{dt}= \underbrace{ 2 \nu \int_0 ^{+ \infty} k^2 T(k,t) dk-4 \nu ^2 \int_0 ^{+ \infty} k^4 E(k,t) dk}_{-C_{\varepsilon 2} \varepsilon^2/K}+\frac{\dot \nu}{\nu} \varepsilon. \label{eq:E}
	\end{align}
\end{subequations}

One recognizes the two first terms on the right hand side of Eq.~(\ref{eq:E}) as the production of dissipation by vortex stretching, here defined from the spectral energy transfer $T(k,t)$, and the palinstrophy  destruction term expressing the damping of velocity gradients by viscosity. Both terms, individually scaling as $\RE ^{1/2}$, are often closed together as $-C_{\varepsilon 2}\,\varepsilon^2/K$ in the classical $K$-$\varepsilon$ model. The viscosity being time dependent, an additional term $\varepsilon \dot \nu /\nu$, which we can refer as the rapid viscosity term, appears in the equation for the dissipation \cite{Coleman1991}. The configurations with small Froude number $\FRO$ thus correspond to the dominance of the rapid viscosity term leading, after integrating Eq.~(\ref{eq:E}), to $\varepsilon \sim \nu$. This explains why the rapid viscosity phases are accompanied with a sudden growth of the dissipation as observed in Figure~\ref{fig:ss}. This aspect is clearly distinct from the enstrophy blow-up phenomenon, often observed at the beginning of simulations, and which is due to the energy transfer towards the smaller scales.   

\begin{figure}
	\begin{center}
		\includegraphics[width=0.7\textwidth]{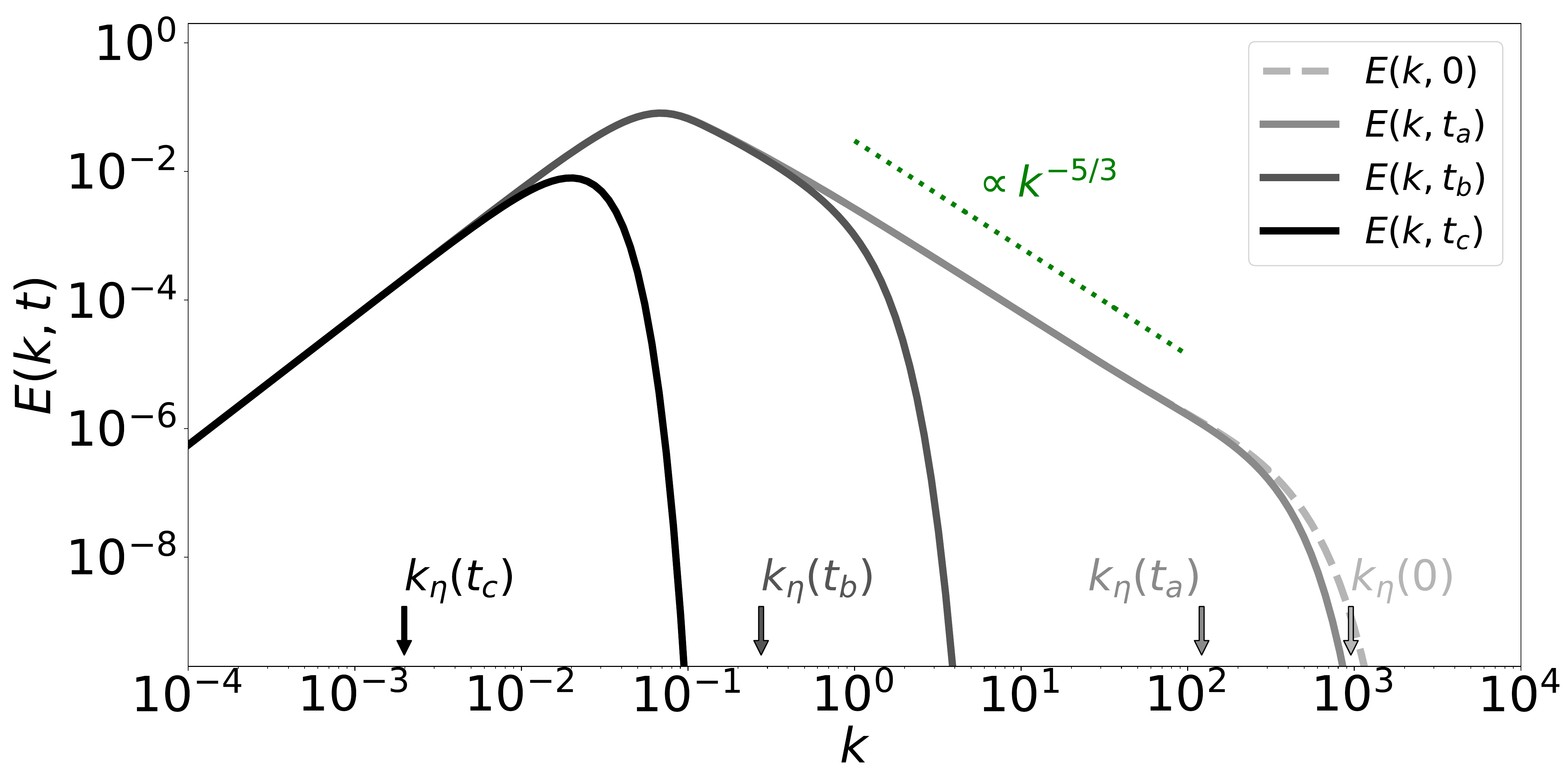}
		\caption{Energy spectra at different times corresponding to the simulation shown in Figure~\ref{fig:ss} with $\theta=2$, initial $\RE_0 \approx 10^6$ and $\FRO_0 = 1.2 \times 10^{-5}$. The different times, $t_{a-c}$, are also indicated in Figure~\ref{fig:ss}. We also place the corresponding inverse Kolmogorov scales defined as $k_\eta=\left(\dfrac{\varepsilon}{\nu^3}\right)^{1/4}$.\label{fig:spec}}
	\end{center}
\end{figure}

It is further useful to analyze how the rapid viscosity regime influences the evolution of turbulent energy spectra. In Figure~\ref{fig:spec}, we show the spectra corresponding to the simulation in the rapid viscosity regime already presented in Figure~\ref{fig:ss} ({\it e.g.} with $\FRO_0 =1.2 \times 10^{-5}$).
At the same times, the inverse Komolgorov scales defined as $k_\eta=\left(\dfrac{\varepsilon}{\nu^3}\right)^{1/4}$
are also plotted. This quantity is often associated with the viscous subrange of a spectrum in spectral equilibrium, as illustrated in the Figure~\ref{fig:spec} at $t=0$. However, as the flow enters the rapid viscous phase, the viscous subrange and the inverse Kolmogorov scale decreasing as $k_\eta \sim \nu ^{-1/2} $ decouple. This expresses the unbalance between the non linear transfers and the viscous dissipative effects. Furthermore, it explains why the kinetic energy at the beginning of the rapid viscous phase is constant leading to the trajectories $\FRO \sim \RE ^{-\dfrac{\theta+1}{2 \theta}}$ already observed in Figure~\ref{fig:evol}.  

Noticeably, the rapid viscosity phases last as long as the Froude number remains small. When the flow exits the rapid viscosity phase, around $\FRO \sim 1$, depending on the value of the Reynolds number, it can enter a self-similar turbulent or viscous regime. We now discuss how these regime changes can be challenging in term of RANS modeling.      

\subsection{The renormalized dissipation rate coefficient}

RANS modeling requires to relate the turbulent dissipation to the large scale properties of the flow. As already discussed in introduction, this is provided by the renormalized dissipation rate coefficient $C_\varepsilon$. In this section, we investigate by the mean of EDQNM simulations how this parameter evolves along the different regimes encountered during the the compression.

\begin{figure}
	\begin{center}
		\includegraphics[width=0.7\textwidth]{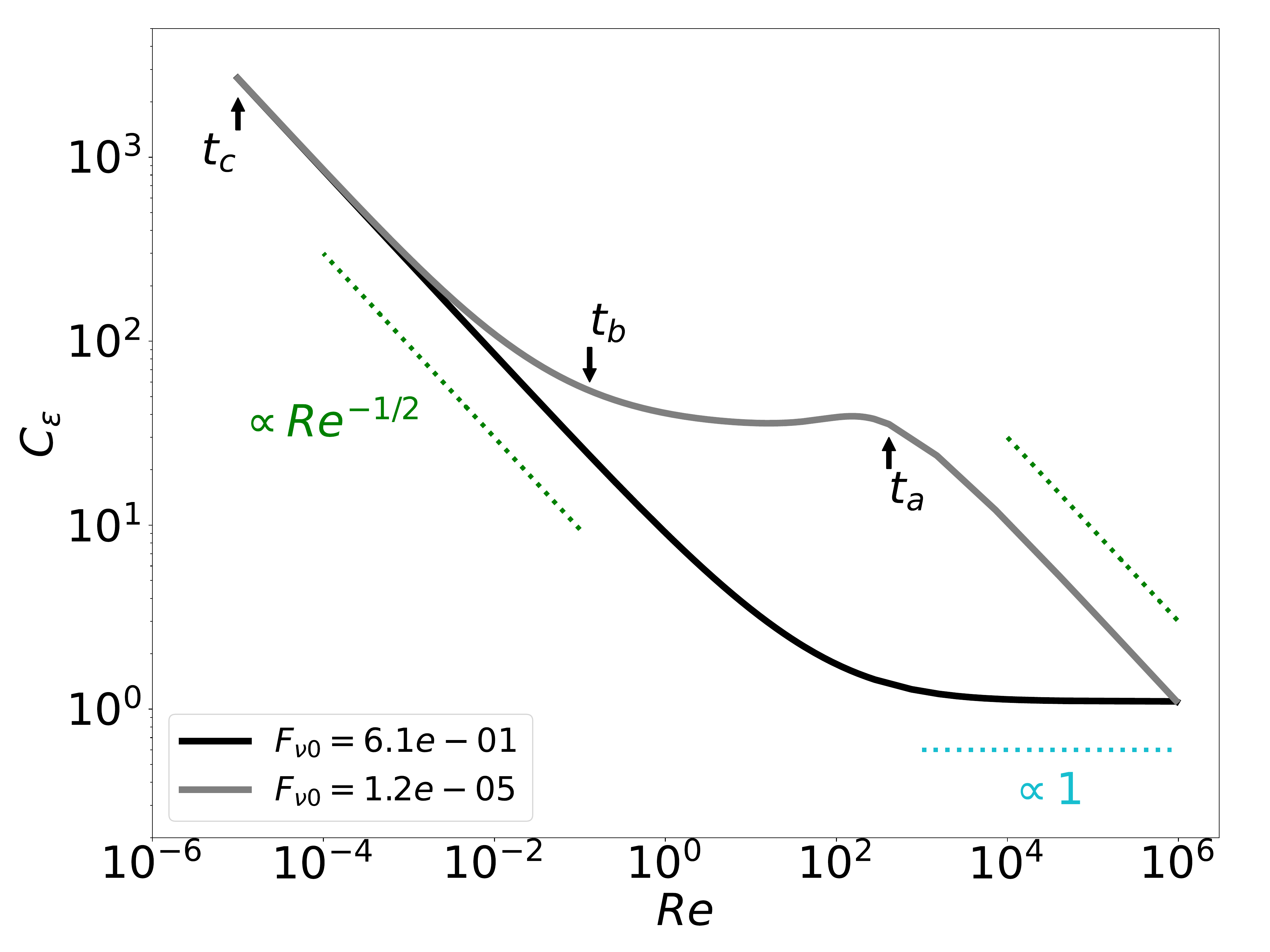}
		\caption{Renormalized dissipation rate as a function of the Reynolds number for different EDQNM simulations initially at $\RE _0 \sim 10^6$. The two curves correspond to the trajectories shown in Figure~\ref{fig:ss} at $\theta=2$ and with the $\FRO_0$ values indicated in the legend. The scaling $\RE^{-1/2}$ corresponding to the dotted green lines shows the self-similar viscous and rapid viscosity regimes. At high Reynolds number, the plateau corresponding to a spectral equilibrium is also plotted. \label{fig:ce}}
	\end{center}
\end{figure}

In homogeneous isotropic turbulence, the longitudinal integral scale $\mathcal L$ and the rms velocity $\mathcal U$ are usually obtained by \cite{Pope2000}
\begin{align}
\mathcal L=\dfrac{3 \pi}{4} \frac{\int_0 ^{+\infty} \frac{E(k)}{k} dk}{\int_0 ^{+\infty}E(k) dk}, \ \text{and} \
\mathcal U=\left( \frac{2}{3} K \right)^{1/2}, \label{eq:def}
\end{align}
which allows the evaluation of the dissipation rate from Eq.~(\ref{eq:C}). Based on this definition, we thus present in Figure~\ref{fig:ce} the dependence of $C_ \varepsilon$ on the Reynolds number $\RE$ for various EDQNM simulations.

The renormalized dissipation rate coefficient is expected to take a constant value at high Reynolds number corresponding to Kolmogorov spectra with a balance between the energy transfer and the dissipation. However, as shown in the Figure~\ref{fig:ce}, the simulations exhibit a strong dependence of the coefficient $C_\varepsilon$ to the Reynolds number. This is particularly true in the low $\RE$ regions where a scaling $\sim \RE^{-1/2}$ is found, associated to the viscous self-similar regime. It is indeed well-known that modeling a turbulent flow experiencing a relaminarization phase necessitates to introduce a Reynolds number dependence in the coefficients. Various models has already been proposed such as \cite{Loshe1994} to express this dependence.

Yet, we observe in the Figure~\ref{fig:ce} that the function $C_\varepsilon (\RE)$ is not universal in particular when the flow enters a rapid viscosity regime (also evidenced by a $\sim \RE^{-1/2}$ scaling). 
In order to account for this crucial aspect, we explore in the next section various strategies in order to derive a model able to reproduce the data from the EDQNM simulations.

\section{Modeling a turbulent plasma under compression}

In this part, we propose to derive a two-equations model either using the classical approach based on self-similar solutions or using modern machine learning techniques allowing the full use of the EDQNM data.  


\subsection{Classical modeling}

Here, we first model turbulence under compression using the two-equations model proposed by \citet{Perot2006} hereafter denoted as PBK. In a second part, we extend the model to rapid viscosity variations by introducing a dependence of its coefficients on the Froude number.

\subsubsection{The PBK Model}

The PBK model \cite{Perot2006} was initially built to reproduce the dissipation in turbulent flows at moderate or small Reynolds numbers close to the walls. Although the model does not account for strong temporal viscosity variations, it is worthy to assess its ability to reproduce the EDQNM data.  

The PBK model expresses the dynamics of the turbulent kinetic energy $K$ and the wavenumber $\lambda$ corresponding to the energy containing eddies. More precisely, the state variable $\lambda$ seeks at reproducing $1/\mathcal L$. In that respect, it differs from $K$-$\varepsilon$ models in which a Reynolds number dependence of $C_{\varepsilon 2}$ is introduced to reproduce the low Reynolds numbers regimes \cite{Jones1972, Durbin1991, Hanjalic1998}. The idea followed by the PBK model is to separate the viscous and turbulent contributions mimicking the Lin equation for the turbulent spectrum. This allows to write the system of equations:     

\begin{subequations}
	\begin{align}
	&\dot K=-\left(\alpha_\nu \nu \lambda^2+ \alpha_T \lambda K^{1/2} \right) K, \label{eq:pbk1}\\
	&\dot \lambda=-\left(\beta_\nu \nu \lambda^2 + \beta_T \lambda K^{1/2} \right) \lambda. \label{eq:pbk2}
	\end{align}
\end{subequations}
In Eqs.~(\ref{eq:pbk1}, \ref{eq:pbk2}), the model constants $\alpha_{\nu, T}$, $\beta_{\nu, T}$ can be constrained to recover the self-similar variable viscosity solutions presented in Sec.~\ref{sec:selfsimilar} as also detailed in \cite{Perot2006}.  
Combining this, the PBK model reduces to
\begin{subequations}
	\begin{align}
	&\dot K=- \frac{1}{\tau}K, \label{eq:pbkb1}\\
	&\dot \lambda=- \frac{1}{(s+1)\tau}\lambda, \ \text{with} \ 1/\tau=\alpha_\nu \nu \lambda ^2+\alpha_T K^{1/2} \lambda,\label{eq:pbkb2}
	\end{align}
\end{subequations}
where only the two constants $\alpha_{\nu, T}$ need to be adjusted. In practice, $\alpha_T$ is determined to recover the value $C_\varepsilon$ at large $\RE$. The constant $\alpha_\nu$ is set to match the transition from the turbulent to the laminar regime and we use the relationship $\alpha_\nu=15 \alpha_T$ already proposed by \cite{Perot2006}. This calibration also accounts for the large scale distribution of energy which corresponds to $s=2$ for our data.    

Noticeably, the PBK model can be put in a $K$-$\varepsilon$ form. Indeed, Eq.~(\ref{eq:pbkb1}) shows that $\lambda$ can be expressed as a function of $\varepsilon$ by finding the positive root of the second order polynomial:
\begin{equation}
\lambda (\varepsilon)=\frac{\sqrt{\alpha_T^2 K ^3+4 \varepsilon \alpha_\nu K \nu}-\alpha_T K^{3/2}}{2 \alpha_\nu K \nu}.\label{eq:pbklambda}
\end{equation}
The equation for $\varepsilon$ is simply obtained by taking the time derivative of Eq.~(\ref{eq:pbkb1}) and using Eqs.~(\ref{eq:pbkb2}) and (\ref{eq:pbklambda}) to eliminate $\lambda$. Its final form leads to a very  lengthy expression for the coefficient $C_{\varepsilon 2}$ which would have been difficult to infer without the intermediate $K$-$\lambda$ variables suggested by PBK.

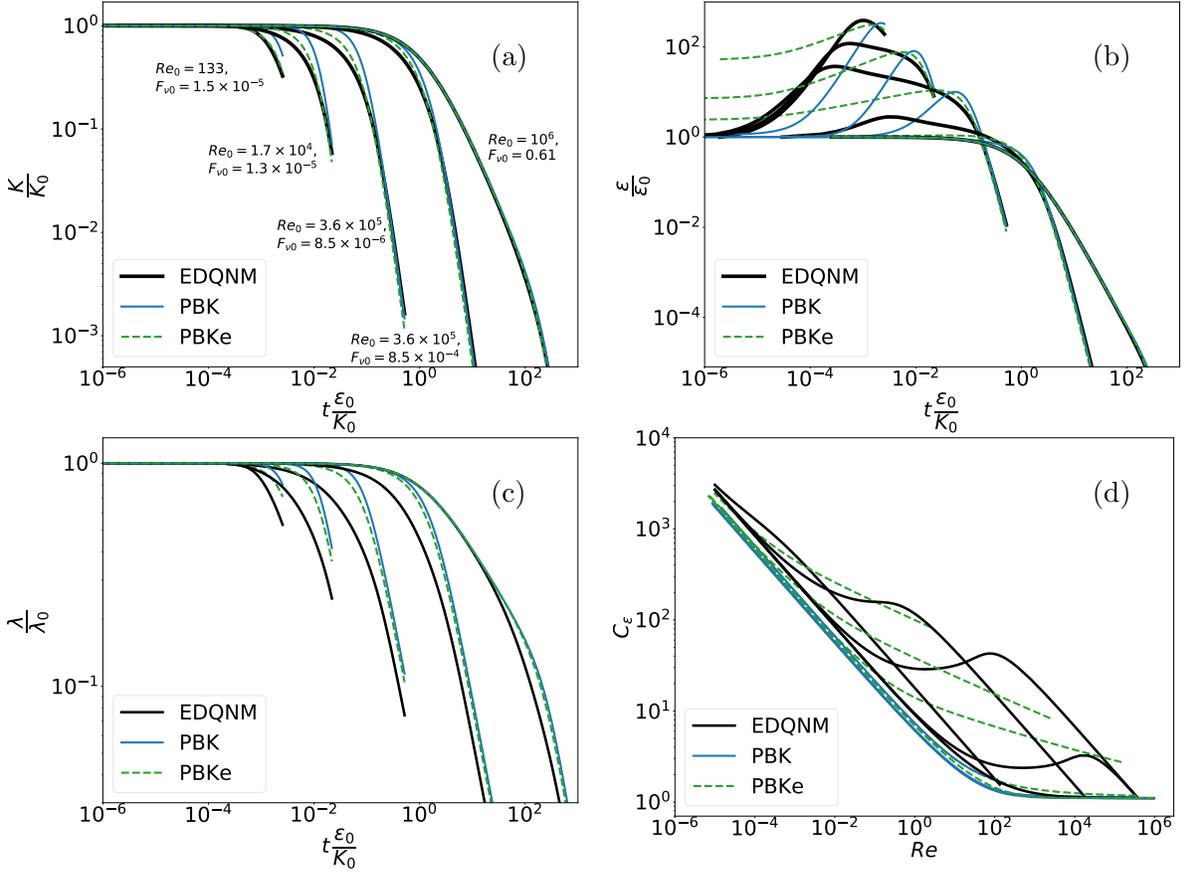
\begin{figure}
	\begin{center}
		\setlength{\unitlength}{1mm}
		\begin{picture}(100,100)
		\put(-30,58){\includegraphics[width=0.48\textwidth]{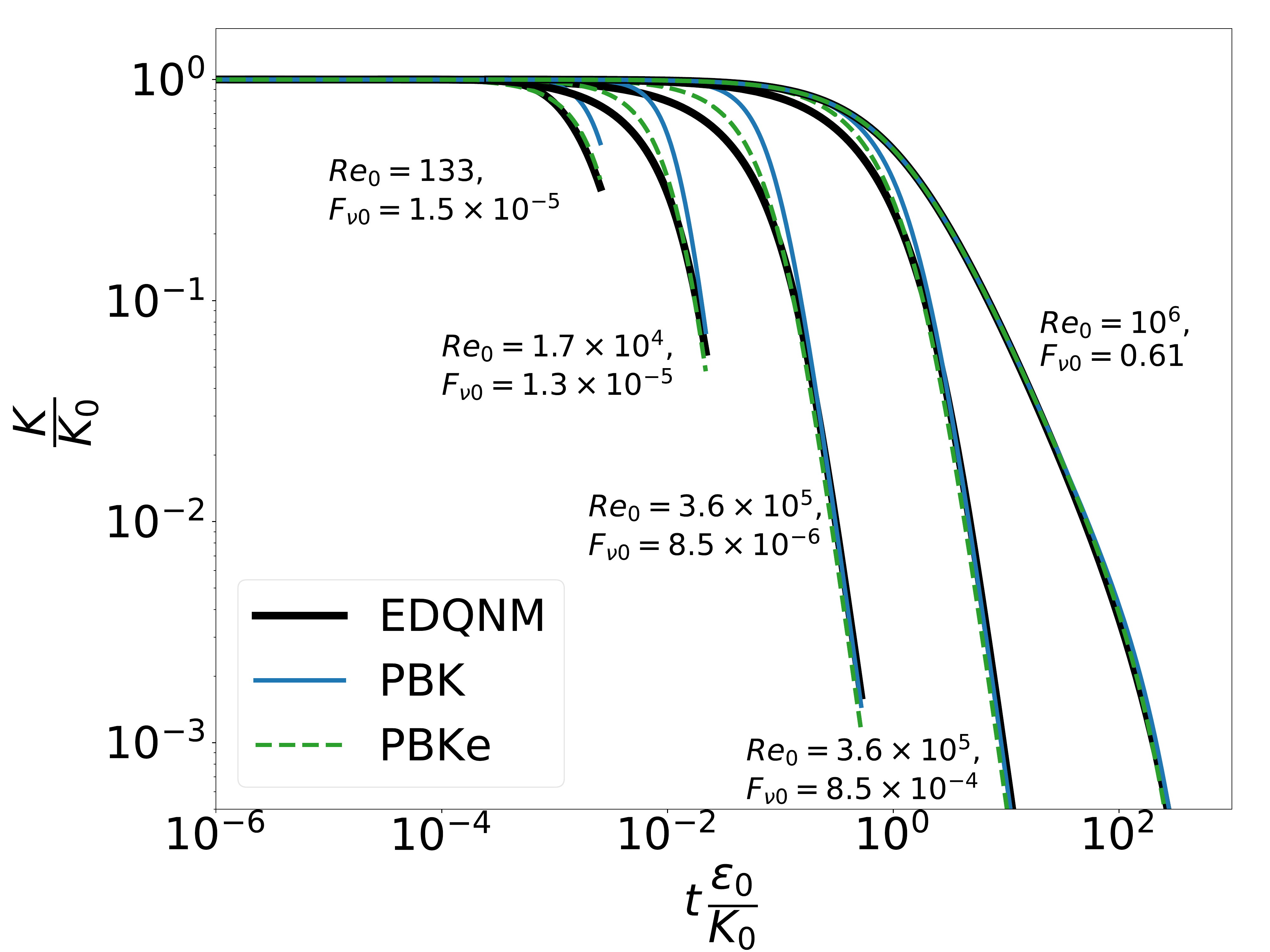}}
		\put(50,58){\includegraphics[width=0.48\textwidth]{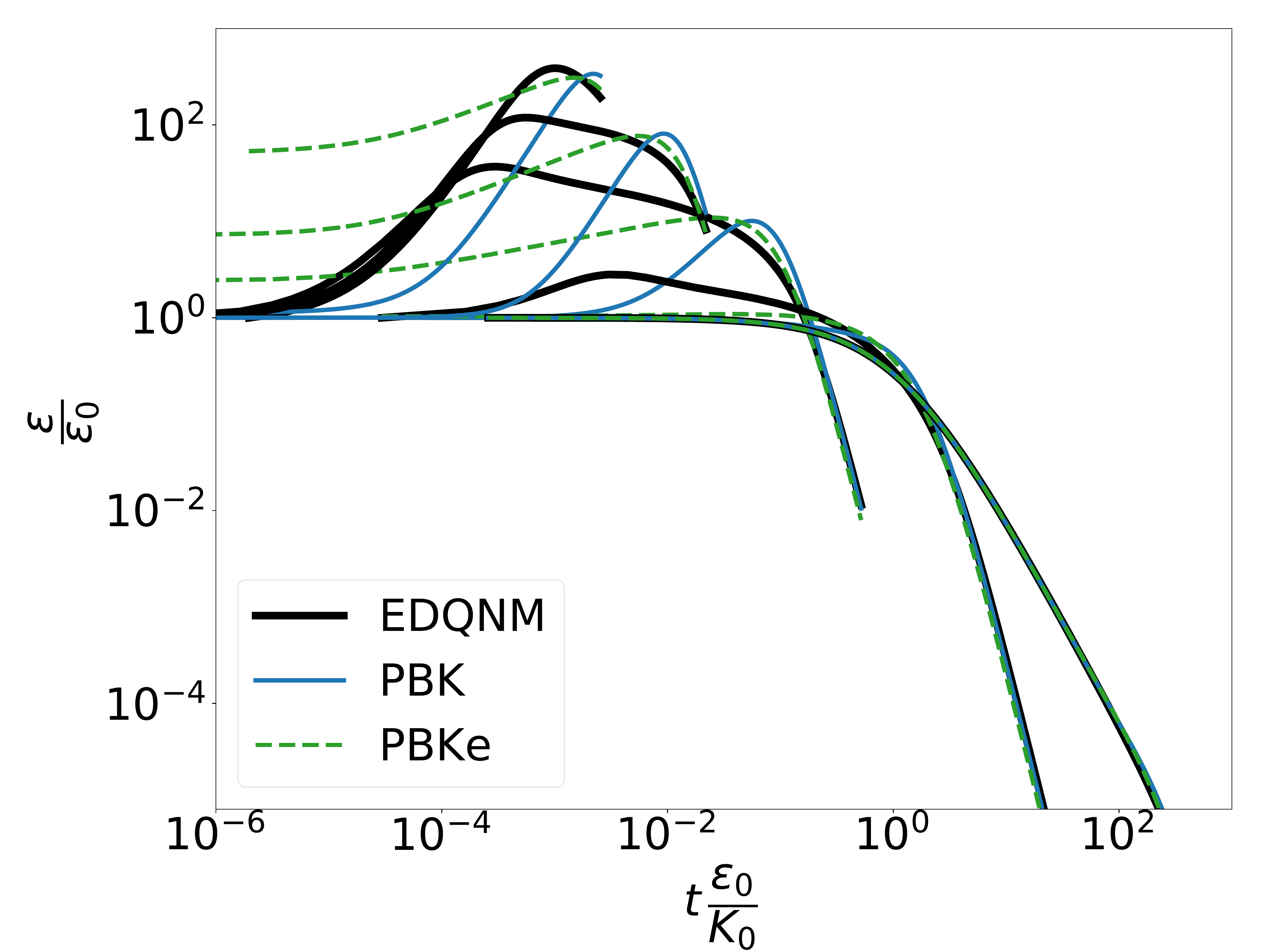}}
		\put(-30,0){\includegraphics[width=0.48\textwidth]{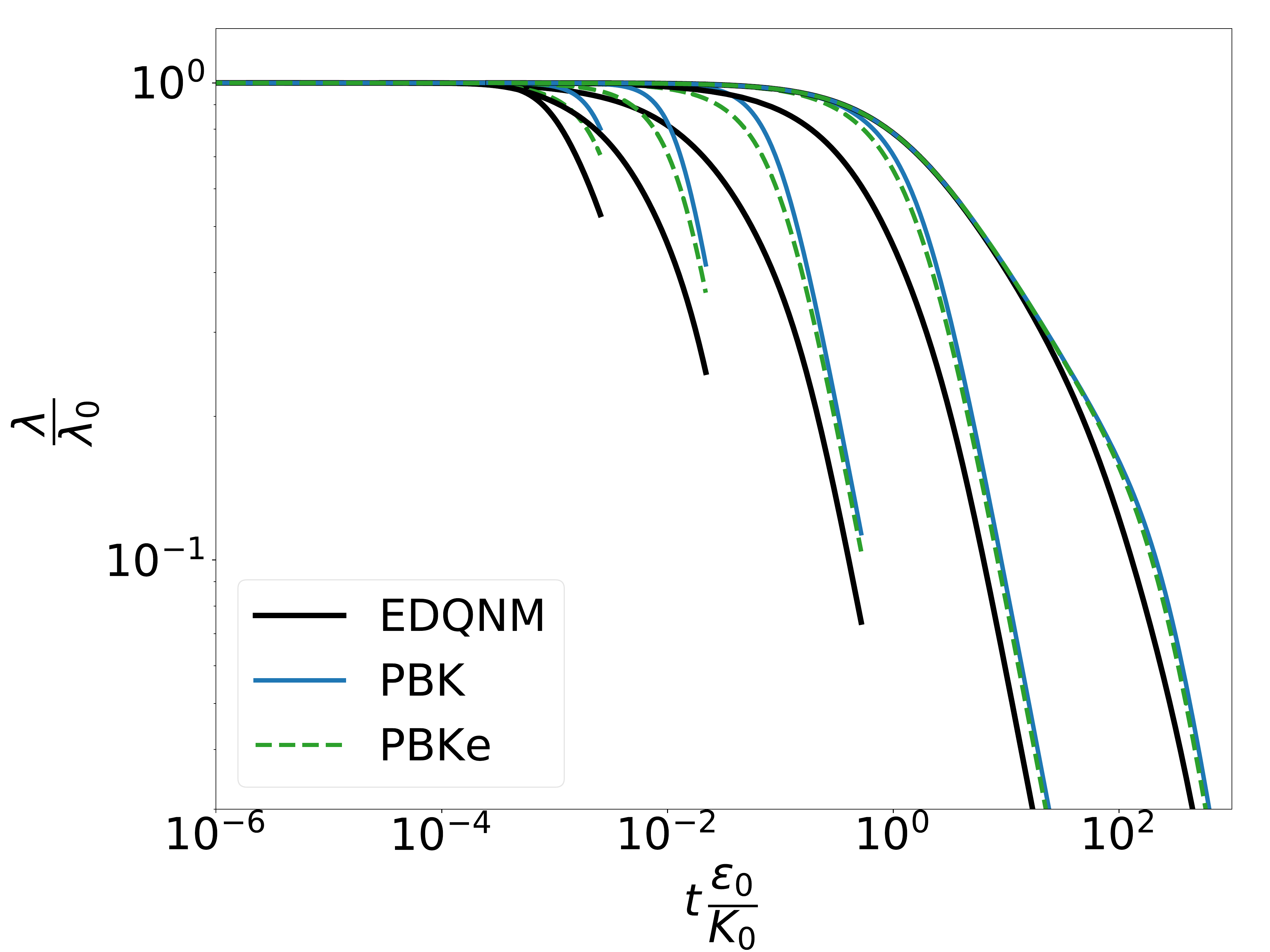}}
		\put(50,0){\includegraphics[width=0.48\textwidth]{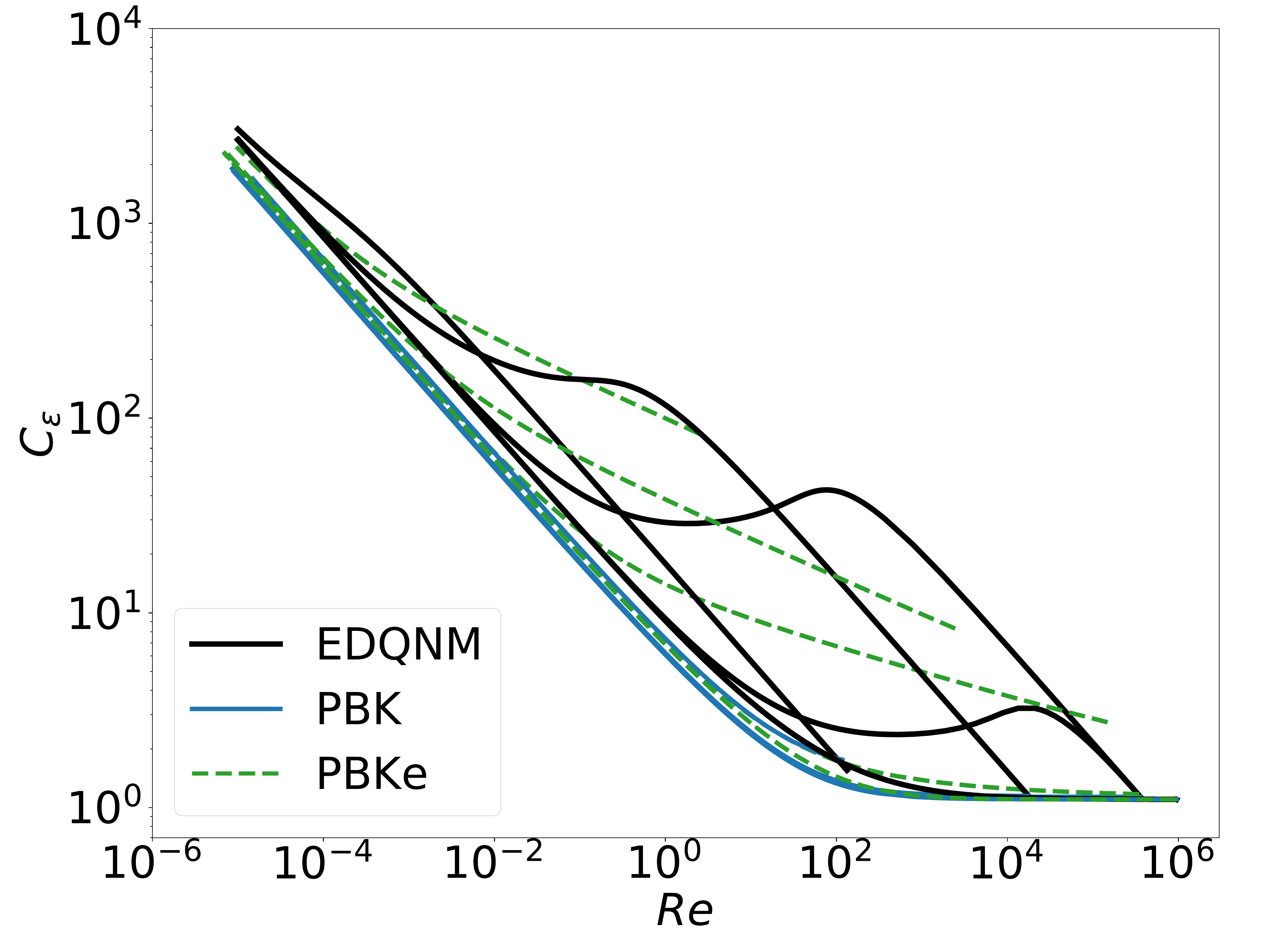}}
		\put(35,107){(a)}
		\put(115,107){(b)}
		\put(35,49){(c)}
		\put(115,49){(d)}
		\end{picture}
		\caption{Comparisons of model PBK and PBKe for representative EDQNM trajectories with respectively: (a) The kinetic energy $K$, (b) the dissipation $\varepsilon$,  (c) the energy containing eddy wavenumber $\lambda$ and (d) the renormalized disipation rate $C_\varepsilon$. The initial Reynolds $\RE _0$ and Froude  $\FRO _0$ numbers for the EDQNM cases are indicated in the Figure (a). \label{fig:compP}}
	\end{center}
\end{figure}
In Figure~\ref{fig:compP}, we present the comparison of the PBK model against representative EDQNM simulations. The configurations without a rapid viscous phase (not too small initial $\FRO _0$) are very well reproduced by PBK. In particular, the turbulent and viscous self-similar phases are correctly captured with the standard  calibration (see Figure~\ref{fig:compP}a). We note a small delay for the evolution of $\lambda$ explaining that although the kinetic energy is correctly predicted, $\lambda$ is slightly overestimated after the laminarization of the flow (see Figure~\ref{fig:compP}c). This effect is also visible with a shift on $C_\varepsilon$ at low Reynolds number between the EDQNM and the PBK curves (see Figure~\ref{fig:compP}d). However and as somehow anticipated, the PBK model exhibits discrepancies with the EDQNM results principally during the rapid viscous phases. In Figure~\ref{fig:compP}b for instance, we observe that the dissipation growth is not sufficiently fast in the PBK model explaining the overestimation of kinetic energy during the rapid regime.    

We now detail how the model can be simply extended to the rapid viscous phases. 

\subsubsection{Extending the PBK model to the rapid viscosity regime}  

The dissipation evolution in a rapid viscous phase is well-known and given by $\dot \varepsilon =\varepsilon \dot \nu / \nu $ as shown by Eq.~(\ref{eq:E}). This suggests to constrain the PBK model in order to recover the rapid solution when the Froude number is small. One can show that adding a supplementary term in the $\lambda$ equation allows this constraint to be verified. 
However, although we ensure the correct initial evolution of $\varepsilon$, the correct values of the plateau are not guaranteed for all the initial conditions. We have not obtained a fully satisfactory model using this method, and we thus try a different approach.   

In order to extend the PBK model to the rapid viscous phases, we propose to introduce a simple dependence of the coefficient $\alpha_\nu$ to the Froude number as 
\begin{equation}
\alpha_{\nu}^{PBKe}=\alpha_{\nu}^{PBK}+\frac{\alpha_e}{\FRO}. \label{eq:pbke}
\end{equation}
The basic idea expressed by Eq.~(\ref{eq:pbke}) is to enhance the dissipation during the rapid viscous phases, {\it i.e $\FRO \ll 1$}. The modified model, hereafter labeled as PBKe, recovers the PBK model when the viscosity variations are negligible. An additional constant $\alpha_e$ is introduced in Eq.~(\ref{eq:pbke}) which is calibrated against the EDQNM data. At first sight, 
a dependence of $\alpha_\nu$ to $\varepsilon$ can be a problem as Eq.~(\ref{eq:pbk1}) becomes implicit. However, one can readily verify that  Eq.~(\ref{eq:pbke}) leads to a second order equation for $\varepsilon$ which can be solved directly.

The results for the PBKe model with calibration $\alpha_e=4$ are also shown in Figure~\ref{fig:compP}.
While the extended model keeps the good properties of PBK in the self-similar regime, a significant improvement is observed during the rapid viscosity phases. Noticeably, the closure for $\varepsilon$ also depends on $\dot \nu$ in PBKe explaining why the trajectories in Figure~\ref{fig:compP}b start at different levels. Here the new constant in PBKe is adjusted in order to recover the different plateau values in the EDQNM simulations.   
Although very satisfactory, the model does not perfectly reproduce the data. Therefore we test the ability of a data-driven approach to obtain better RANS closures for this problem.  

\subsection{A data-driven approach}

Machine learning for turbulence modeling have proven an efficient tool when a large amount of data is available \cite{Duraisamy2019,Brenner2019,brunton_machine_2020}. Many methods have been developed and tested to this aim, among them neural networks (NN) are popular \cite{Ling2016,wu_physics-informed_2018,karniadakis_physics-informed_2021}. To gain confidence in the model,  it is often preferable to seek for interpretative and parsimonious closures. This is allowed by sparse regression strategies such as SINDy (Sparse Identification of Nonlinear Dynamics) \cite{Brunton2016,Beetham2020} which we apply here to the problem of a compressed turbulent plasma.

\subsubsection{Methodology}

In this section, we describe how to adapt the SINDy method in order to derive two-equations models able to reproduce the EDQNM data.
The data are therefore randomly decomposed into a training set for the regression containing 100 EDQNM trajectories (70 \% of the data) and a validation set with the remaining 43 trajectories used for the model selection.

For the regression, we define a target vector of dimension $2 N_d$ as
\begin{equation}
{\bf Y}=
\begin{pmatrix} {\bf Y}_K \\ {\bf Y}_\lambda \end{pmatrix} \label{eq:target}.
\end{equation}
In this expression, ${\bf Y}_K$ and ${\bf Y}_\lambda$ are column vectors of size $N_d$ formed from respectively the renormalized derivative of kinetic energy, $\dot K \nu/K \dot \nu$, and the large eddy wavenumber $\dot \lambda \nu/ \lambda \dot \nu$ (we still use $\lambda=1/\mathcal L$) at various times and from various EDQNM trajectories of the training set. In practice, the dimension of the target vector ${\bf Y}$ is roughly $2 N_d=230000$. For this problem involving different physical variables such as $K$ and $\lambda$, it seems important to consider dimensionless quantities. Furthermore, these quantities vary over several orders of magnitude along the trajectories and re-scaling them is helpful for the regression algorithm.   

In the SINDy methodology, we try to fit the target vector using a basis of candidate functions $\bf{\Theta (X)}$ associated with a weight vector ${\bf \Xi}$ and assembled as $\bf{\Theta (X)}{\bf \Xi}$. Similarly to the target vector, the state vector is defined in a dimensionless form as ${\bf X}=(R,F)$. Here we introduce the pseudo Reynolds and Froude numbers constructed from the model variables as $R=K^{1/2}/(\nu \lambda)$ and $F=\dot \nu/ (\nu^2 \lambda^2)$ as suggested by dimensional analysis.
One element of the function basis $ {\bf \Theta}$ is thus formed from the combination $R^{h_R} F^{h_F}$ evaluated at a given time in an EDQNM trajectory of the training data set.
The exponents introduced for the basis functions are also constrained as $h_F \in [-1 \ 0]$ and $h_R \in[0 \ 1]$. These interval bounds correspond to the two terms of the PBK model which control the turbulent and viscous self-similar regimes. It seems natural to propose this basis as we seek for a smoother transition from the turbulent to the viscous regime and knowing that the PBK model is already satisfying except in the rapid viscous regime.
The exponents $h_F$ and $h_R$ are sampled on these intervals with $\delta h_F =0.25$ and $\delta h_R=0.5$ such that we try to reproduce each terms, $\dot K$ and $\dot \lambda$, on a basis of $N_f=15$ functions. The problem can be cast on a matrix form as :
\begin{equation}
\bf{\Theta (X)}{\bf \Xi}=\begin{pmatrix} {\bf \Theta}_K {\bf (X)} & 0 \\ 0 & {\bf \Theta}_\lambda {\bf (X)} \end{pmatrix}
\begin{pmatrix} {\bf \Xi}_K \\ {\bf \Xi}_\lambda \end{pmatrix} \label{eq:fit},
\end{equation}
with the matrix ${\bf \Theta}$ of dimension $2 N_d \times 2 N_f$, and with  sub-matrices ${\bf \Theta}_K$ and ${\bf \Theta}_\lambda$ (which are here identical of size $N_d \times N_f$) corresponding to the basis functions to reproduce $\dot K$ and $\dot \lambda$, in association with the weight vectors ${\bf \Xi}_K$ and ${\bf \Xi}_\lambda$ (of size $N_f$).

To solve this problem, we apply the sequentially thresholded least squares (STLSQ) algorithm with a ridge regularization, in which the objective function is
\begin{equation}
	\begin{aligned}
		\mathcal L({\bf \Xi})=\| {\bf Y} -{\bf \Theta(X) \Xi} \|_2 ^2+\eta_K \| {\bf \Xi}_K\| ^2_2+\eta_\lambda \| {\bf \Xi}_\lambda \| ^2_2, \\
		&\text{ such that } \ \left|\xi^j_{K,\lambda}\right|\geq\sigma_{K,\lambda} \ \ \forall\, \xi^j_{K,\lambda}\text{ in } \boldsymbol{\Xi}_{K,\lambda}.
	\end{aligned}
\end{equation}
The first two hyperparameters $\eta_K$ and $\eta_\lambda$ control the strength of the ridge regularization. A sub-iteration of this regression procedure consists in finding the solution $\text{argmin}_{{\bf \Xi}}\, \mathcal L({\bf \Xi})$, which can be determined analytically. In order to ensure parsimony, a hard-thresholding process is applied.
The coefficients $\xi_K$ in ${\bf \Xi}_K$ and $\xi_\lambda$ in ${\bf \Xi_\lambda}$ with absolute values lower than given thresholds $\sigma_K$ and $\sigma_\lambda$, and their associated candidates, are removed from the regression problem. The whole procedure is repeated until convergence. 
Finally, the entire sparse regression is performed for several values of the four hyperparameters leading to a great number of combinations ($\eta_K,\eta_\lambda,\sigma_K,\sigma_\lambda$) tried and as many corresponding solutions.

In order to select the best model among all of those, we assess their performance on the EDQNM solutions from the validation data set. For a given trajectory, the model is integrated from the initial conditions. We then compute a mean relative error on $K$ and $\lambda$ as:
\begin{equation}
\epsilon^i_K=\frac{1}{N_p}\Sigma_{p} \left| \frac{K^{EDQNM}-K^{model}}{K^{EDQNM}}\right|, \ \epsilon^i_\lambda=\frac{1}{N_p}\Sigma_{p} \left| \frac{\lambda^{EDQNM}-\lambda^{model}}{\lambda^{EDQNM}}\right|
\end{equation}
where $N_p$ is the numbers of time steps on a given trajectory $i$. We then select only the models in the hyperparameter space verifying that the maximum errors $\epsilon_{K, \lambda}$ among all the trajectories is less than a given accuracy $\mathcal E$, {\it i.e} $\max_i \epsilon^i_{K,\lambda} \le \mathcal E$. On this first selection, we further consider the more parsimonious ones, {\it i.e} those having the smallest number of terms $N$ from the basis function. Lastly, a final model is obtained as the one  having the smallest maximum of relative error over all the trajectories from the pre-selected subset.

\subsubsection{Results}

In this part, we apply the previous methodology to find two-equations models able to reproduce our EDQNM data. 

We have developed our own Python code which has been first tested on data generated by the PBK model. Noticeably, the PBK model was successfully identified with the correct coefficients by SINDy. This is not surprising as the closure terms from PBK are part of the basis functions. Importantly, this test gives us confidence in that we have reached the minimum number of trajectories necessary to find a model.  

\begin{figure}
	\begin{center}
		\setlength{\unitlength}{1mm}
		\begin{picture}(100,100)
		\put(-30,58){\includegraphics[width=\textwidth]{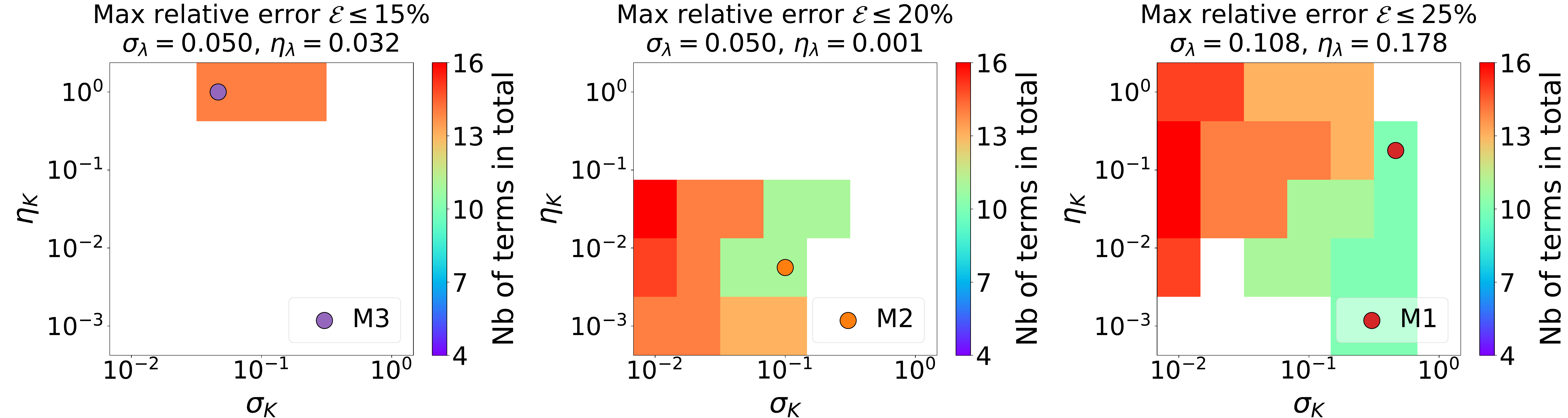}}
		\put(-30,0){\includegraphics[width=\textwidth]{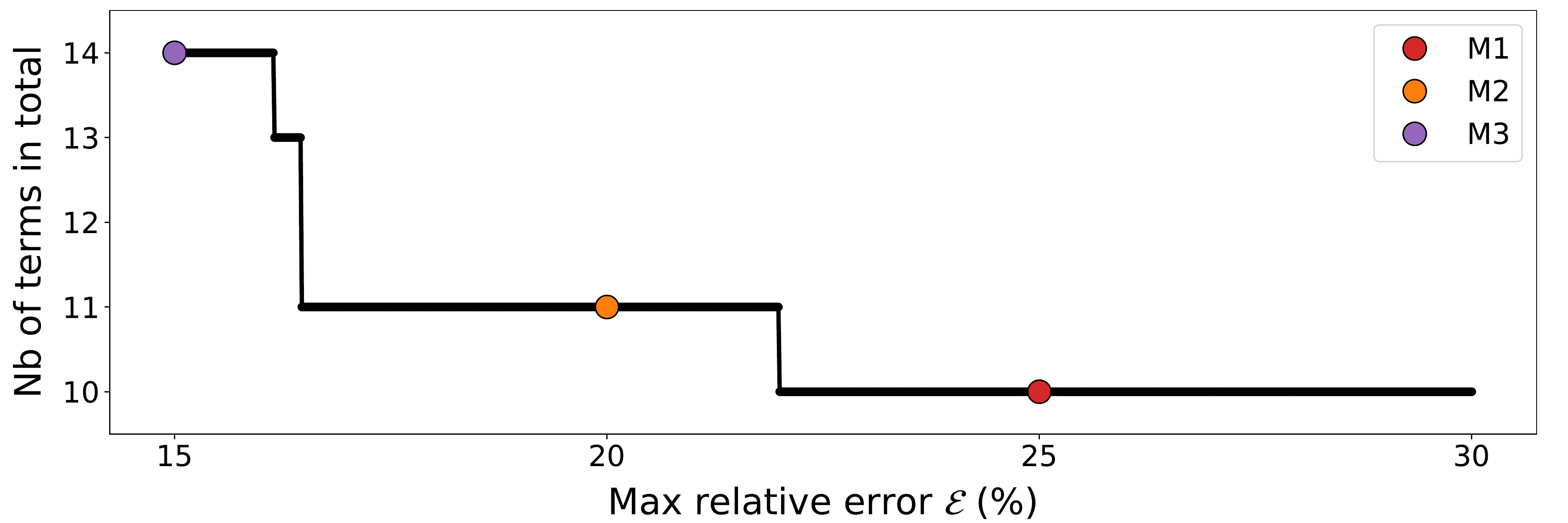}}
		\put(-30,98){(a)}
		\put(-30,55){(b)}
		\end{picture}
		\caption{Model selection procedure in the regression problem applied to the EDQNM data. (a) 2D ($\eta_K$-$\sigma_K$) planes extracted from the four dimensions hyperparameter space showing the number of terms in the regression solution satisfying various maximum relative errors. These planes pass through the best solutions corresponding to model M1, M2 and M3 with maximum relative error $\mathcal E=$ 25, 20 and 15 \% respectively. (b) Number of terms in the selected model as a function of the maximum of relative error allowed. The M1, M2 and M3 models identified in the hyperparameter space are also indicated. 
			\label{fig:selection}}
	\end{center}
\end{figure}

Hence, we apply the method on the EDQNM training data set as shown in the Figure \ref{fig:selection}.
Given a maximum relative error, we find in the hyperparameter space the most parsimonious model satisfying this criterion as shown in the Figure~\ref{fig:selection}a for instance. As expected, more terms in the model are requested for a  better accuracy. This aspect is well illustrated in the Figure~\ref{fig:selection}b showing the decrease of the number of terms required when the accuracy is degraded.
This analysis evidences that the maximum relative error principally comes from the $K$ trajectories ($\epsilon^i_K > \epsilon^i_\lambda$) and also that more terms in the equation for $\lambda$ are needed. We have managed to find parsimonious solutions with 15 \% maximum error. Decreasing further would request a better refinement and extension of the basis function.
We will now consider more specifically the 3 models, M1, M2 and M3, derived for a maximum relative error $\mathcal E$ of 25, 20 and 15 \% respectively. 

\begin{figure}
	\begin{center}
		\setlength{\unitlength}{1mm}
		\begin{picture}(100,100)
		\put(-30,58){\includegraphics[width=0.48\textwidth]{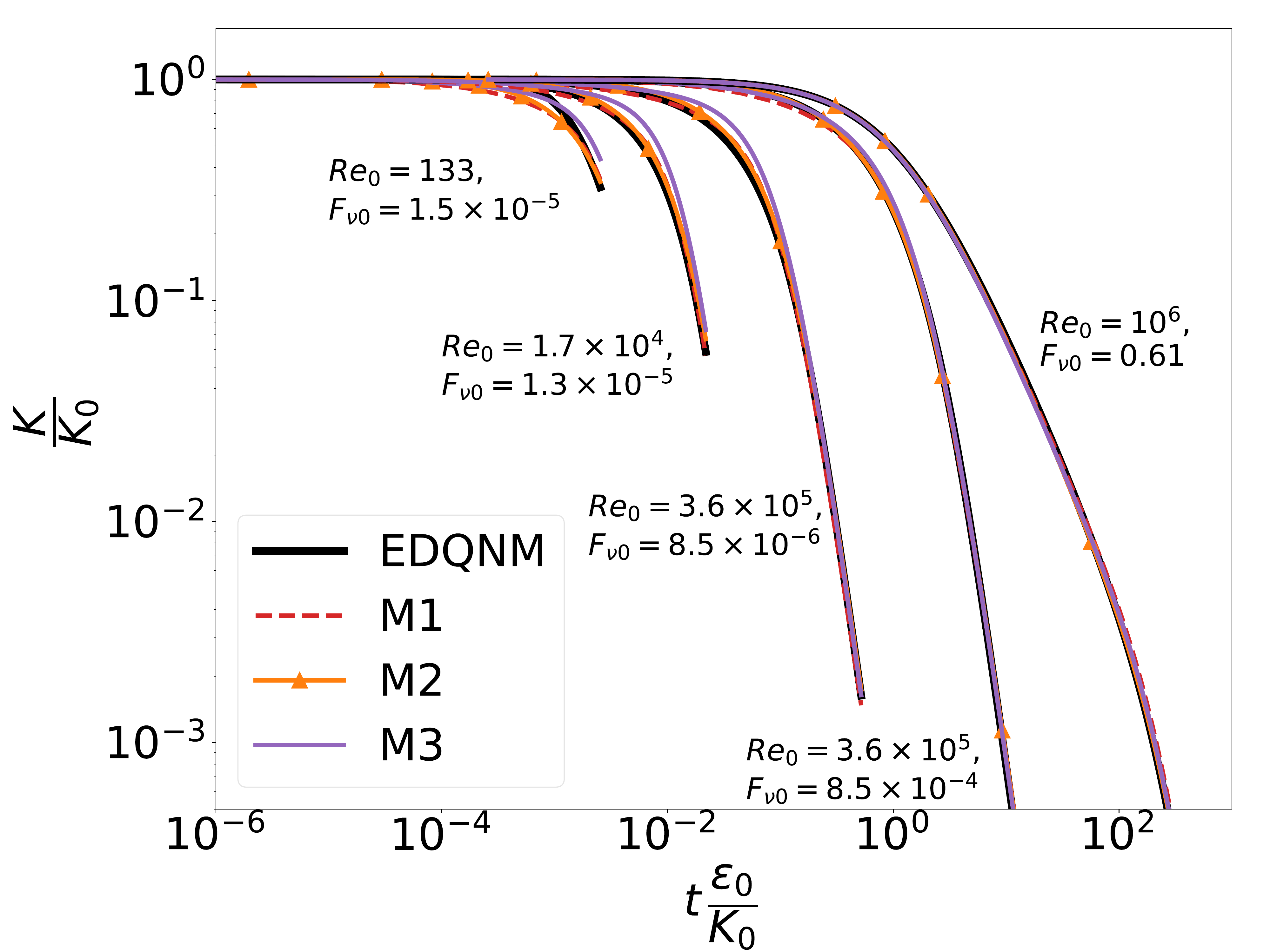}}
		\put(50,58){\includegraphics[width=0.48\textwidth]{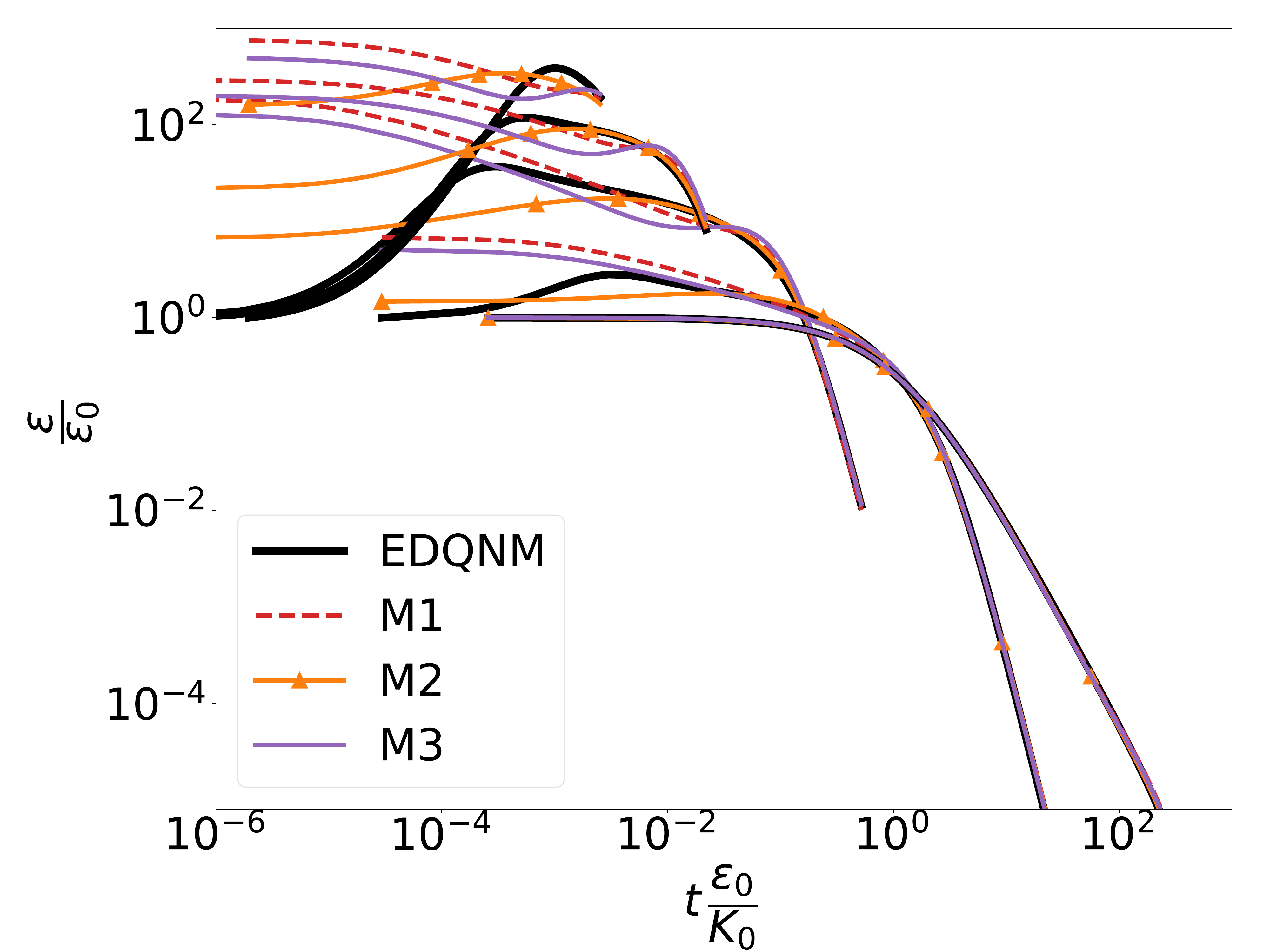}}
		\put(-30,0){\includegraphics[width=0.48\textwidth]{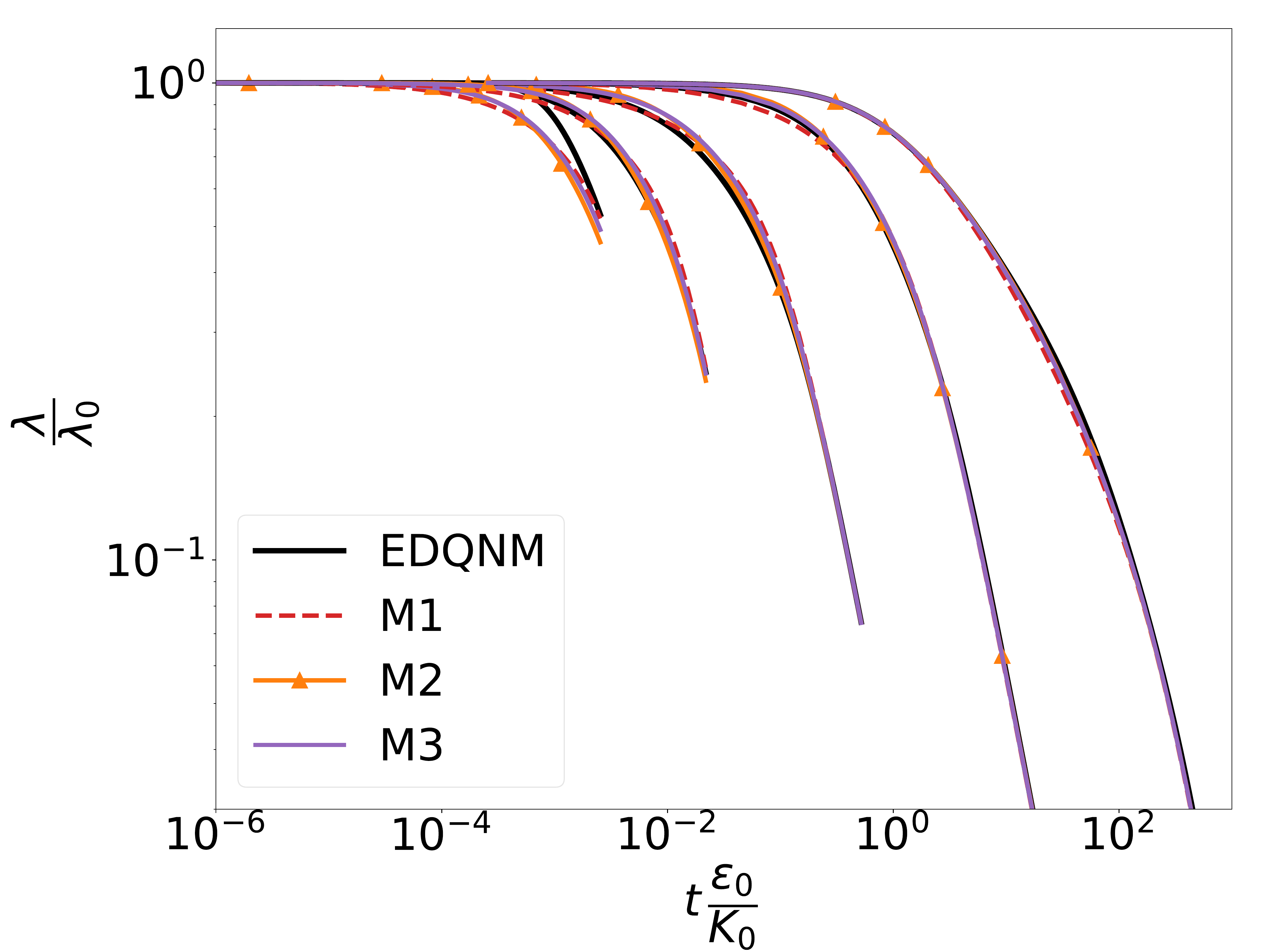}}
		\put(50,0){\includegraphics[width=0.48\textwidth]{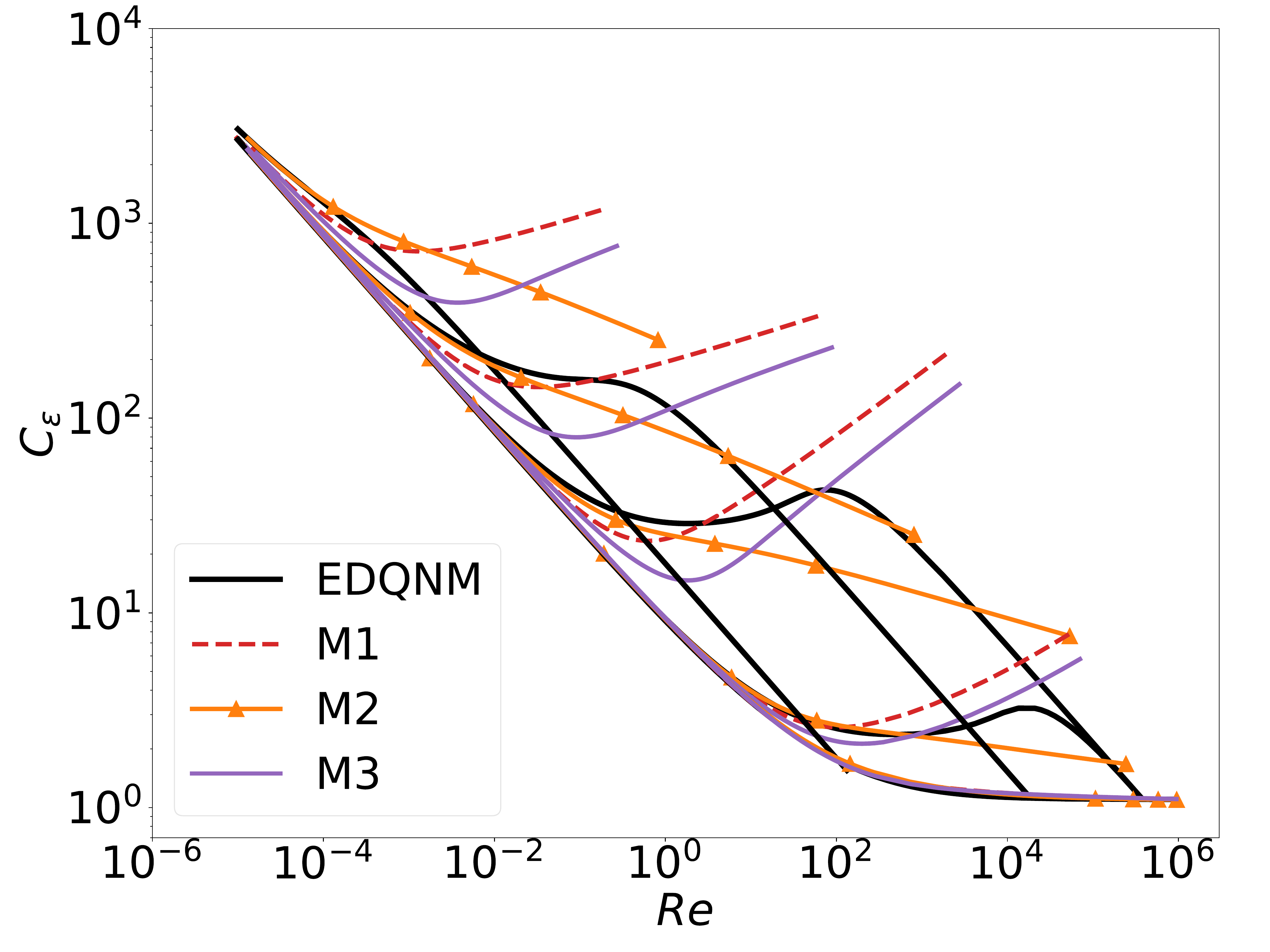}}
		\put(35,107){(a)}
		\put(115,107){(b)}
		\put(35,49){(c)}
		\put(115,49){(d)}
		\end{picture}
		\caption{Comparisons of models M1-3 for representative EDQNM trajectories, with respectively: (a) The kinetic energy $K$, (b) the dissipation $\varepsilon$,  (c) the energy containing eddy wavenumber $\lambda$ and (d) the renormalized disipation rate $C_\varepsilon$. The initial Reynolds $\RE _0$ and Froude  $\FRO _0$ numbers for the EDQNM cases are indicated in the Figure (a). \label{fig:compM}}
	\end{center}
\end{figure}
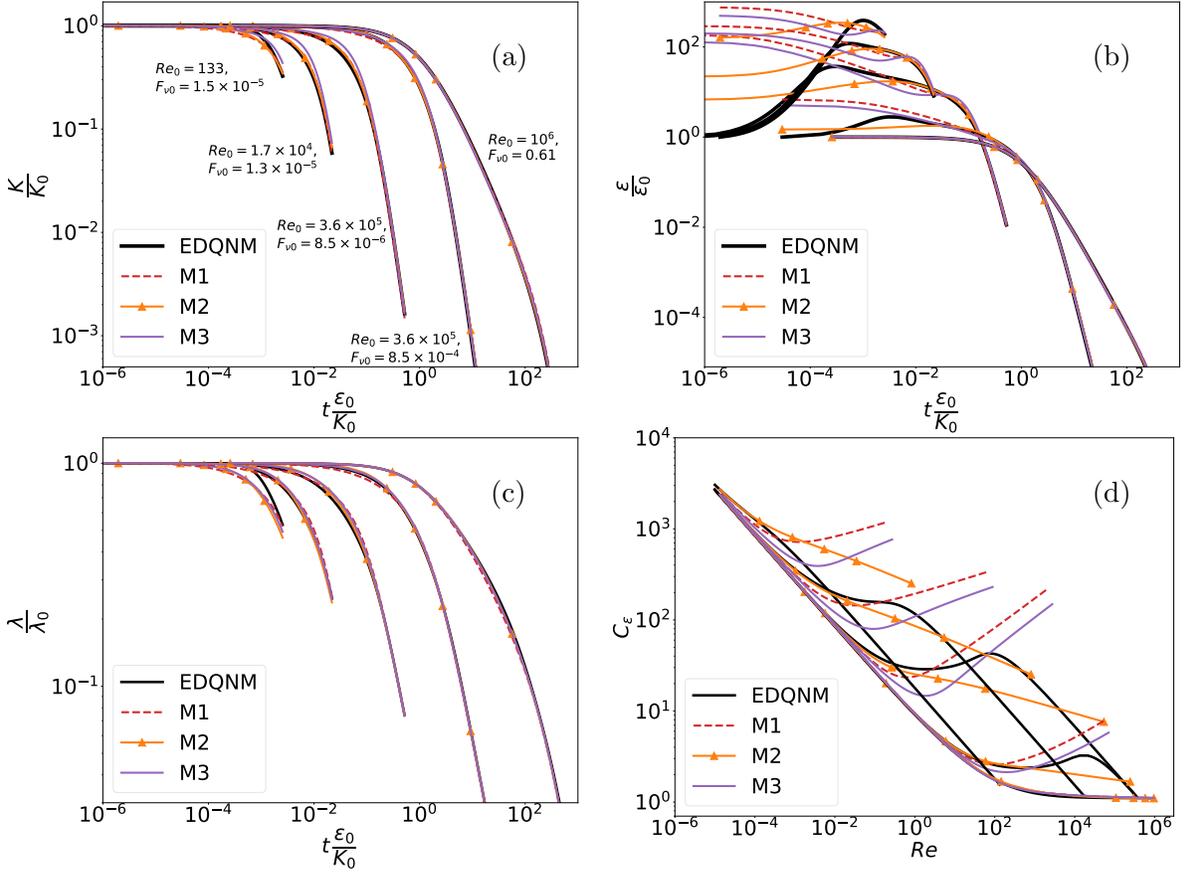

Similarly to the Figure~\ref{fig:compP} comparing the PBK and the EDQMN models, we present in the Figure~\ref{fig:compM} the evolution of kinetic energy (a), dissipation (b), wavenumber (c) and $C_\varepsilon$ (d) for the models M1-3 on the same EDQNM trajectories. Note that these trajectories were not used in the regression procedure nor included in the validation set. The results exhibit a good agreement with EDQNM for all the models M1-3, both on $K$ and $\lambda$, even on trajectories including rapid viscous phases. In addition, one can notice that models M2-3 having a larger number of terms are visibly more accurate than M1. This demonstrates the efficiency of SINDy to derive turbulence closure for compressed plasma.    

\begin{figure}
	\begin{center}
		\setlength{\unitlength}{1mm}
		\begin{picture}(100,100)
		\put(-30,54){\includegraphics[width=0.95\textwidth]{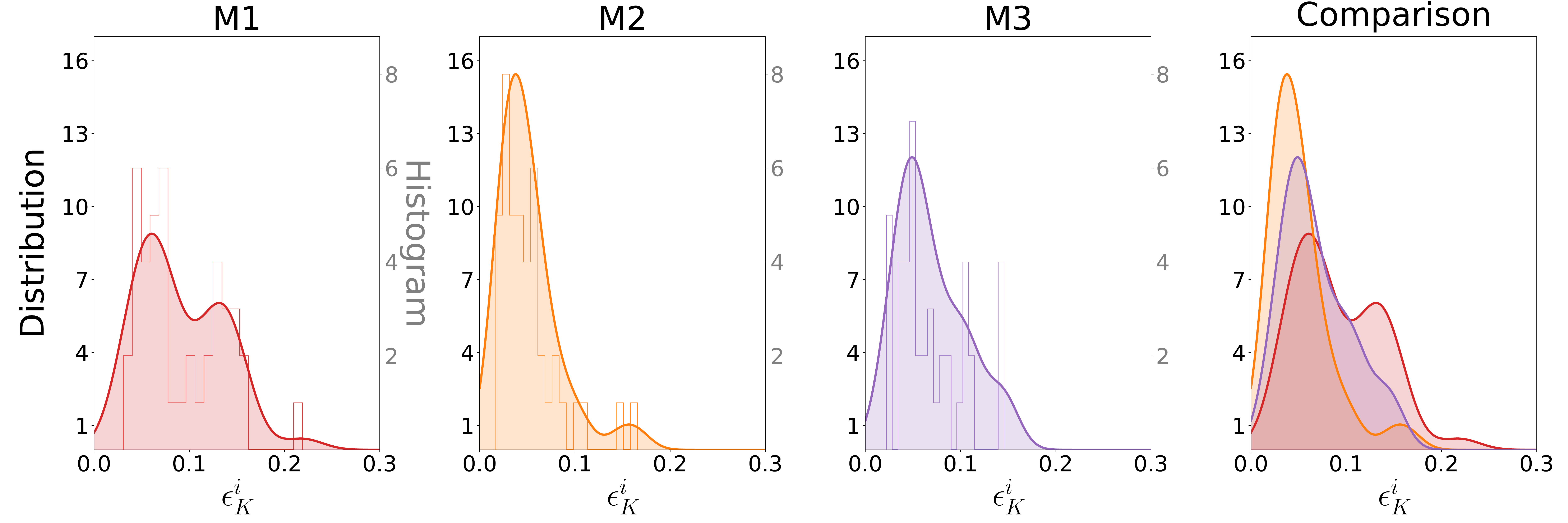}}
		\put(-30,0){\includegraphics[width=0.95\textwidth]{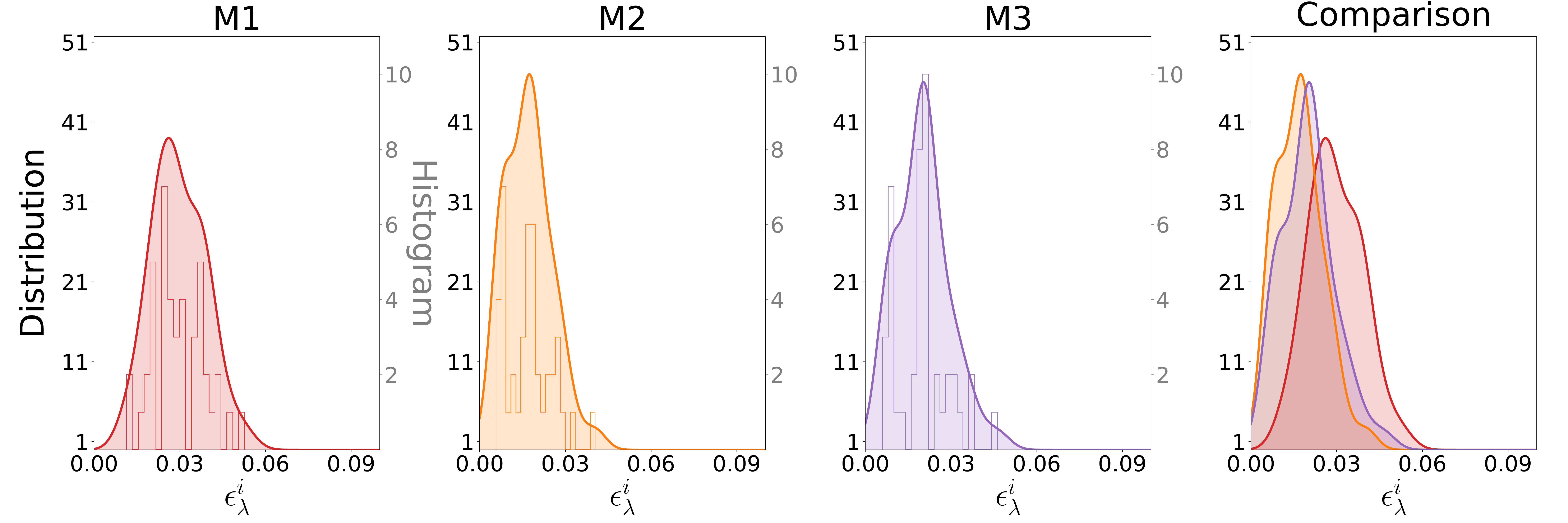}}
		\put(-30,103){(a)}
		\put(-30,50){(b)}
		\end{picture}
		\caption{Distribution of the mean relative errors  $\epsilon$ on $K$ (a) and $\lambda$ (b) for all the trajectories of the validation data set and for the models M1-3. \label{fig:compTOT}}
	\end{center}
\end{figure}

In order to quantify more thoroughly the performance of the models M1-3 derived from the regression procedure, we present in the Figure~\ref{fig:compTOT} the distribution of the mean relative error for each trajectory of the validation data set, both on $K$ (a) and $\lambda$ (b). For all the models, the mean relative error is 
smaller in the $\lambda$ equation than for $K$, $\epsilon_\lambda \le \epsilon _K$, confirming that the models are principally selected from their ability to reproduce the dynamics of $K$. The criterion based on the maximum of mean relative error does not guarantee that the selected model has the lowest mean error. Indeed the M2 model, while having less terms than M3, is nevertheless better in term of mean error as shown by the distributions of the Figure~\ref{fig:compTOT}. It is therefore a good candidate to be evaluated against the PBK and PBKe.    

Therefore, we focus on the M2 model involving 5 and 6 terms ($N=11$) respectively in the equations for $K$ and $\lambda$ given by:
\begin{subequations}
	\begin{equation}
	\dot{K}= -\alpha_1\,\nu\lambda^2 K -\alpha_2\,\lambda K^{3/2} +\alpha_3\,\dot{\nu}^{1/4}\nu^{1/2}\lambda^{3/2} K +\alpha_4\,\dot{\nu}^{1/4}\lambda K^{5/4} - \alpha_5\,\dot{\nu}^{1/2}\lambda K, \label{eq:m2K}
	\end{equation}
	\begin{equation}
	\dot{\lambda}= -\beta_1\,\nu\lambda^3 - \beta_2\,\nu^{1/2}\lambda^{5/2}K^{1/4} -\beta_3\,\lambda^2 K^{1/2} +\beta_4\,\dot{\nu}^{1/4}\nu^{1/2}\lambda^{5/2} +\beta_5\,\dot{\nu}^{1/4}\lambda^2 K^{1/4} -\beta_6\,\dot{\nu}^{1/2}\lambda^2.\label{eq:m2LAM}
	\end{equation}
\end{subequations}

\begin{center} 
\begin{table}
\begin{tabular}{ccccccc}
Coefficients  i  & $1$ &$2$&$3$&$4$&$5$&$6$ \\ \hline
$\alpha_i$& $36.374$&$0.598$&$19.932$& $0.204$ & $6.525$ &\\
$\beta _i$ &$22.441$&$1.067$ &$0.207$&$20.267$&$0.501$ &$6.586$
\\
\hline
\end{tabular}
\caption{Coefficients for the M2 model obtained from the regression procedure.
\label{tab:M2}}
\end{table}
\end{center}

Here, the model coefficients $\alpha_i$ and $\beta_i$ introduced in Eqs. (\ref{eq:m2K}) and (\ref{eq:m2LAM}) are provided in the Table~\ref{tab:M2}. 
One can notice that the terms on the right hand side of the M2 equations have different signs. In order to keep $K$ and $\lambda$ positive (realizability condition), it is sufficient to ensure that $\dot K/K < 0$ and $\dot \lambda /\lambda <0$. It is possible to show easily that this condition is verified for the M2 model. For instance, the right hand sides of the equations can be turned into second order polynomials for $\lambda^{1/2}$.

\begin{figure}
	\begin{center}
		\setlength{\unitlength}{1mm}
		\begin{picture}(100,100)
		\put(-30,50){\includegraphics[width=0.95\textwidth]{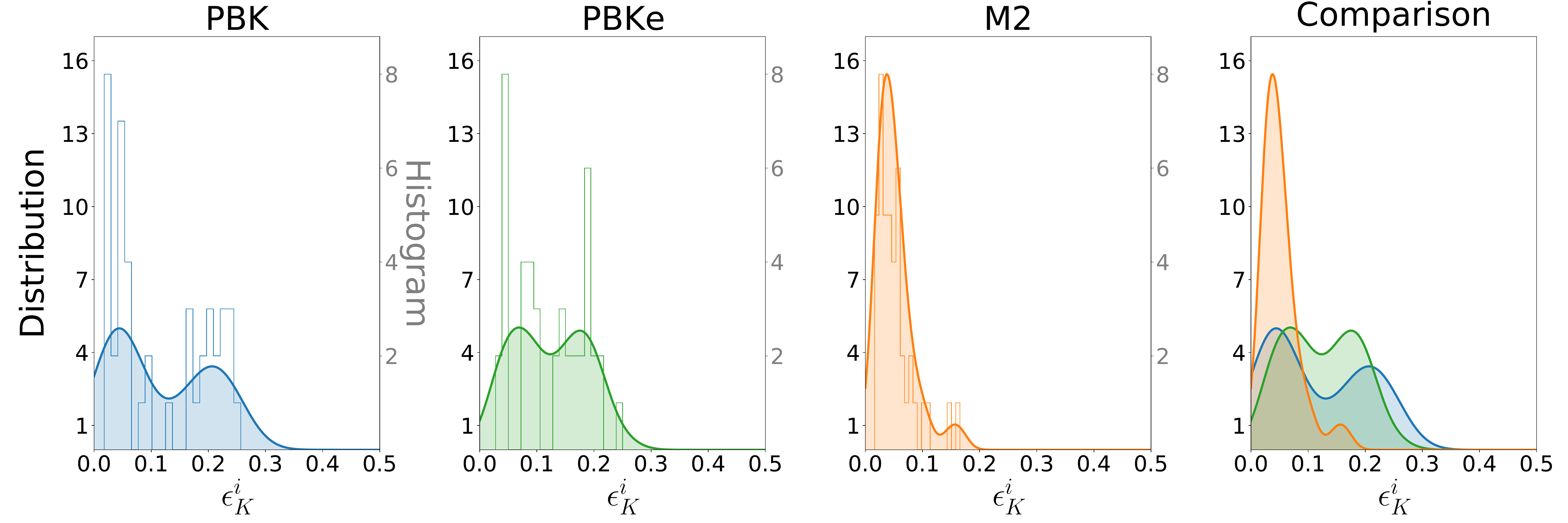}}
		\put(-30,-5){\includegraphics[width=0.95\textwidth]{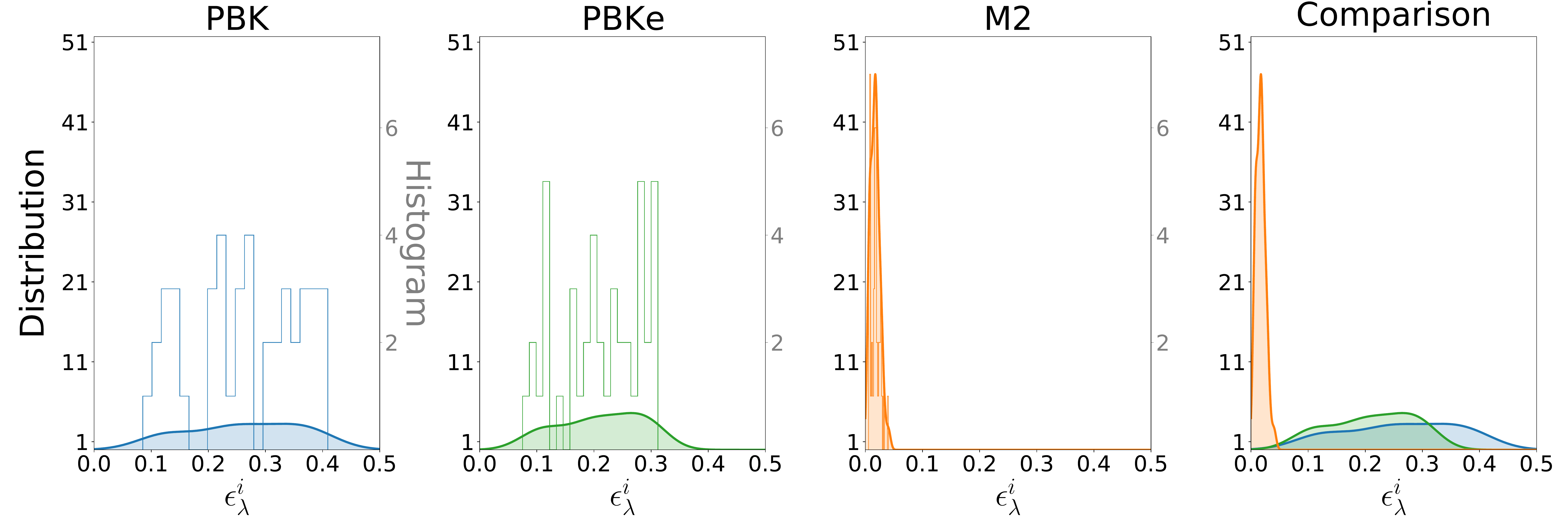}}
		\put(-30,100){(a)}
		\put(-30,45){(b)}
		\end{picture}
		\caption{Distribution of the mean relative errors on $K$ (a) and $\lambda$ (b) for all the trajectories of the validation data set and for the models PBK, PBKe and M2. \label{fig:compbTOT}}
	\end{center}
\end{figure}

We now compare the two-equations models derived from the classical and the machine learning approaches. Similarly to the Figure~\ref{fig:compTOT}, the Figure~\ref{fig:compbTOT} presents the distributions of the mean relative errors for the PBK, PBKe and M2 models. 
Clearly, the M2 model significantly improves the accuracy of the predictions, both in terms of mean or maximum error. Perhaps the most significant gain is on the $\lambda$ equation where M2 performs much better than PBK and PBKe exhibiting flat and extended distributions. In homogeneous isotropic turbulence under compression, the prediction of the integral scale is not a necessity to correctly capture the kinetic energy. It becomes crucial in more complex problems though, as to close the turbulent diffusion terms of an inhomogeneous mixing layer.  

\begin{figure}
	\begin{center}
		\setlength{\unitlength}{1mm}
		\begin{picture}(100,60)
		\put(-30,0){\includegraphics[width=0.45\textwidth]{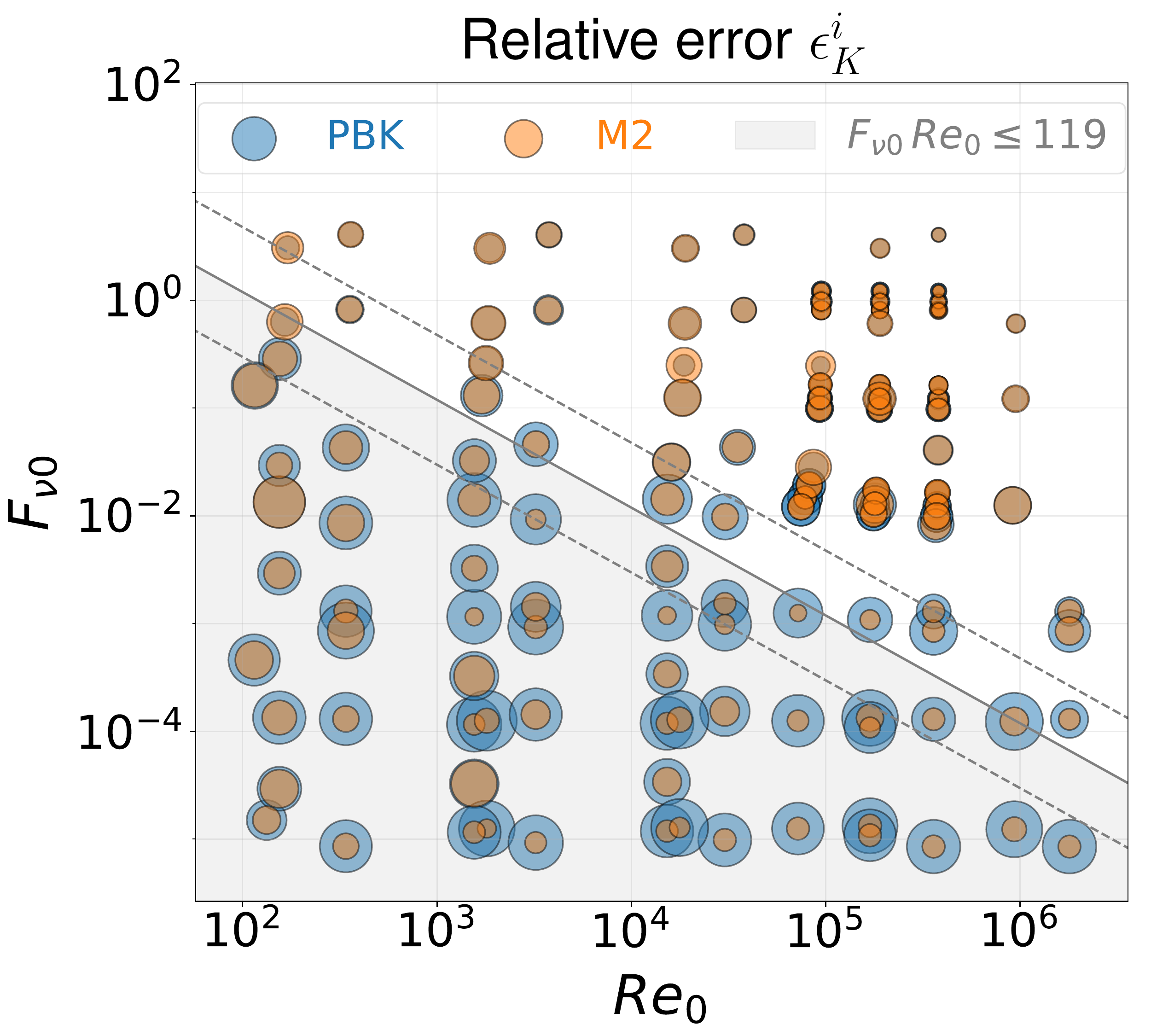}}
		\put(50,0){\includegraphics[width=0.45\textwidth]{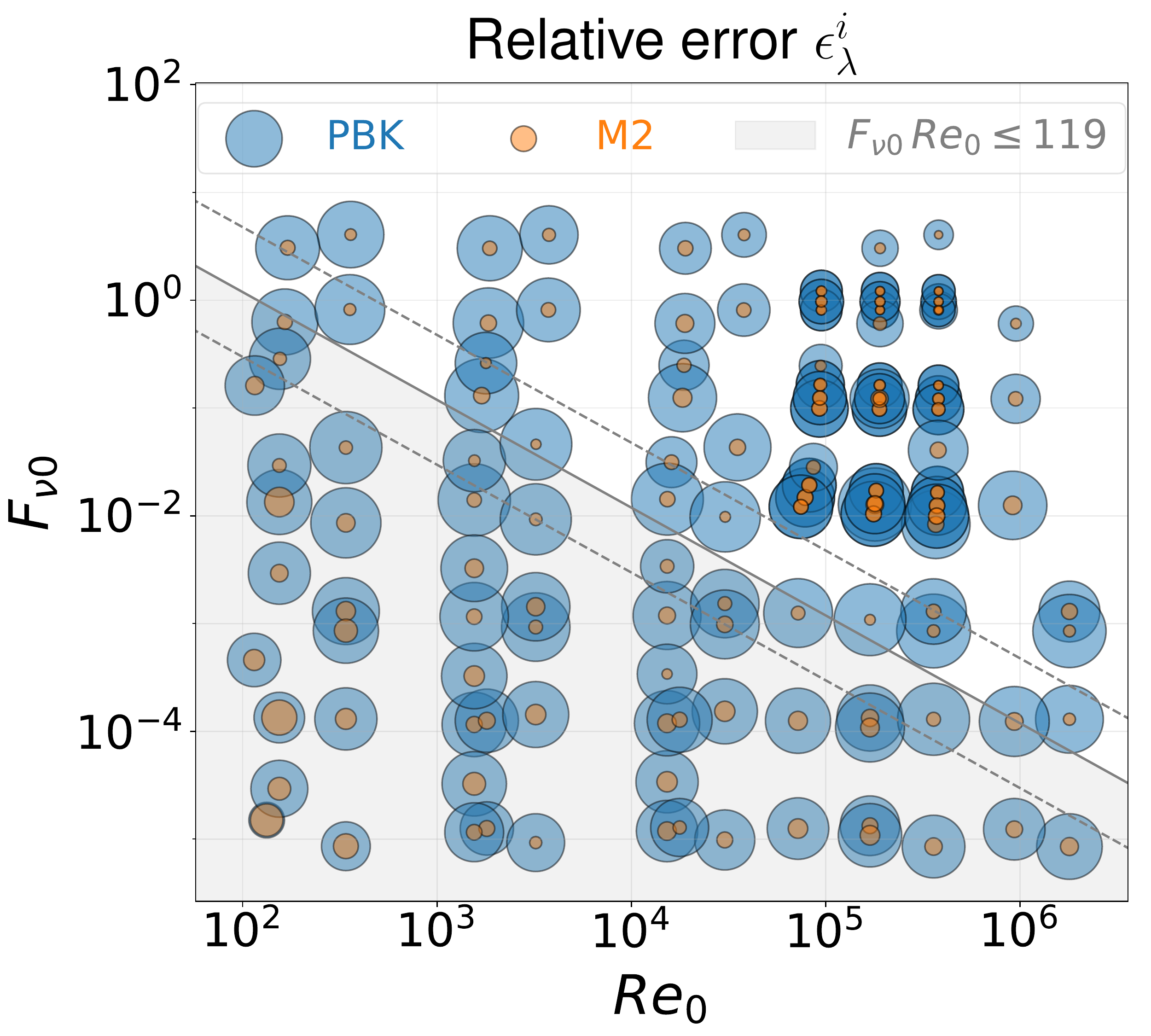}}
		\put(-30,50){(a)}
		\put(50,50){(b)}
		\end{picture}
		\caption{Mean relative error $\epsilon$, on $K$ (a) and $\lambda$ (b), for the models PBK and M2 on the trajectories parametrized by the initial Reynolds and Froude numbers. The region $ \FRO \RE \le 119 $ corresponding to the rapid viscous regime derived from the M2 model is also indicated. \label{fig:emap}}
	\end{center}
\end{figure}

In order to get insight of the PBK and M2 models, we show in the Figure~\ref{fig:emap} the mean relative errors $\epsilon$ for both K (a) and $\lambda$ (b) for the trajectories parameterized by the initial Froude and Reynolds numbers. 
One can see the better performance of the M2 model in particular in the small $\RE$ and $\FRO$ regions. Turning back to Eq.~\ref{eq:m2K} for the M2 model, the balance between the $\alpha_2$ and $\alpha_5$ terms indicates the transition to the rapid viscous regime. This criterion can also be written as $ \FRO \RE \le (\alpha_5/\alpha_2)^2= 119 $ using the constants values. Therefore, the M2 model improves sensitively PBK in this specific region corresponding to the rapid viscous regime as shown by the Figure. 

\section{CONCLUSION} 

In this work, we have considered the problem of an homogeneous turbulent plasma under compression and how to predict its dynamics from simple two-equations turbulence models.

Depending on the plasma viscosity growth, a relaminarization of the flow can occur. Different self-similar turbulent or viscous regimes have been identified, extending \cite{Viciconte2018} to various viscosity power-law  growth exponents. As our analysis does not suppose the confinement of the integral scale, we find that the viscosity growth exponent should be slightly larger than the predictions of \cite{Davidovits2017} for a sudden viscous dissipation to develop. Interestingly, when the viscosity growth is faster than the turbulent eddy turnover time, a rapid viscous regime develops as identified by \cite{Coleman1991}, characterized by the decoupling between the dissipation at small scales and the large scale properties of the flow. This phenomenology has been confirmed on numerous EDQNM simulations for varying initial Froude and Reynolds numbers. 

We then assess the possibility of simple turbulence models to account for the relaminarization and the rapid viscous effects. To this aim, we have tested the $K$-$\lambda$ PBK model \cite{Perot2006} to reproduce our EDQNM data. After an adequate calibration, the PBK model has been able to reproduce fairly well the dynamics of the compressed turbulent plasma except in the rapid viscous regimes. We thus have proposed an extension of the PBK model by introducing a dependence of the coefficient to the Froude number, significantly improving the results. Noticeably, this latter model can be easily turned into a classical $K$-$\varepsilon$ model commonly used by engineers. Thanks to the large amount of data produced by the EDQNM simulations, we have tested a machine learning modeling strategy relying on the SINDy method. We have been able to obtain parsimonious and explainable two-equations models able to reproduce the EDQNM data even in the rapid viscous phases. The success of SINDy relies also on the choice of the function basis which has been guided by dimensional analysis and the PBK model structure.
The benefit of this promising strategy is not only to accurately reproduce the dynamics of kinetic energy but also properly capture the integral scale which is important in the closure of inhomogeneous diffusion terms among others.

\bibliographystyle{apsrevlong}
\bibliography{bibvisc}

\begin{thebibliography}{41}
\expandafter\ifx\csname natexlab\endcsname\relax\def\natexlab#1{#1}\fi
\expandafter\ifx\csname bibnamefont\endcsname\relax
  \def\bibnamefont#1{#1}\fi
\expandafter\ifx\csname bibfnamefont\endcsname\relax
  \def\bibfnamefont#1{#1}\fi
\expandafter\ifx\csname citenamefont\endcsname\relax
  \def\citenamefont#1{#1}\fi
\expandafter\ifx\csname url\endcsname\relax
  \def\url#1{\texttt{#1}}\fi
\expandafter\ifx\csname urlprefix\endcsname\relax\def\urlprefix{URL }\fi
\providecommand{\bibinfo}[2]{#2}
\providecommand{\eprint}[2][]{\url{#2}}

\bibitem[{\citenamefont{Viciconte et~al.}(2018)\citenamefont{Viciconte,
  Gr{\'e}a, and Godeferd}}]{Viciconte2018}
\bibinfo{author}{\bibfnamefont{G.}~\bibnamefont{Viciconte}},
  \bibinfo{author}{\bibfnamefont{B.-J.} \bibnamefont{Gr{\'e}a}},
  \bibnamefont{and} \bibinfo{author}{\bibfnamefont{F.~S.}
  \bibnamefont{Godeferd}}, {``}\bibinfo{title}{Self-similar regimes of
  turbulence in weakly coupled plasmas under compression},{''}
  \bibinfo{journal}{Physical Review E} \textbf{\bibinfo{volume}{97}},
  \bibinfo{pages}{023201} (\bibinfo{year}{2018}).

\bibitem[{\citenamefont{Coleman and Mansour}(1991)}]{Coleman1991}
\bibinfo{author}{\bibfnamefont{G.~N.} \bibnamefont{Coleman}} \bibnamefont{and}
  \bibinfo{author}{\bibfnamefont{N.~N.} \bibnamefont{Mansour}},
  {``}\bibinfo{title}{Modeling the rapid spherical compression of isotropic
  turbulence},{''} \bibinfo{journal}{Physics of Fluids A {:} Fluid Dynamics}
  \textbf{\bibinfo{volume}{3}}, \bibinfo{pages}{2255} (\bibinfo{year}{1991}).

\bibitem[{\citenamefont{Clark et~al.}(2015)\citenamefont{Clark, Marinak, Weber,
  Eder, Haan, Hammel, Hinkel, Jones, Milovich, Patel et~al.}}]{Clark2015}
\bibinfo{author}{\bibfnamefont{D.~S.} \bibnamefont{Clark}},
  \bibinfo{author}{\bibfnamefont{M.~M.} \bibnamefont{Marinak}},
  \bibinfo{author}{\bibfnamefont{C.~R.} \bibnamefont{Weber}},
  \bibinfo{author}{\bibfnamefont{D.~C.} \bibnamefont{Eder}},
  \bibinfo{author}{\bibfnamefont{S.~W.} \bibnamefont{Haan}},
  \bibinfo{author}{\bibfnamefont{B.~A.} \bibnamefont{Hammel}},
  \bibinfo{author}{\bibfnamefont{D.~E.} \bibnamefont{Hinkel}},
  \bibinfo{author}{\bibfnamefont{O.~S.} \bibnamefont{Jones}},
  \bibinfo{author}{\bibfnamefont{J.~L.} \bibnamefont{Milovich}},
  \bibinfo{author}{\bibfnamefont{P.~K.} \bibnamefont{Patel}},
  \bibnamefont{et~al.}, {``}\bibinfo{title}{Radiation hydrodynamics modeling of
  the highest compression inertial confinement fusion ignition experiment from
  the National Ignition Campaign},{''} \bibinfo{journal}{Physics of Plasmas}
  \textbf{\bibinfo{volume}{22}}, \bibinfo{pages}{022703}
  (\bibinfo{year}{2015}), \eprint{https://doi.org/10.1063/1.4906897},
  \urlprefix\url{https://doi.org/10.1063/1.4906897}.

\bibitem[{\citenamefont{Remington et~al.}(2018)\citenamefont{Remington, Park,
  Casey, Cavallo, Clark, Huntington, Kuranz, Miles, Nagel, Raman
  et~al.}}]{Remington2018}
\bibinfo{author}{\bibfnamefont{B.~A.} \bibnamefont{Remington}},
  \bibinfo{author}{\bibfnamefont{H.-S.} \bibnamefont{Park}},
  \bibinfo{author}{\bibfnamefont{D.~T.} \bibnamefont{Casey}},
  \bibinfo{author}{\bibfnamefont{R.~M.} \bibnamefont{Cavallo}},
  \bibinfo{author}{\bibfnamefont{D.~S.} \bibnamefont{Clark}},
  \bibinfo{author}{\bibfnamefont{C.~M.} \bibnamefont{Huntington}},
  \bibinfo{author}{\bibfnamefont{C.~C.} \bibnamefont{Kuranz}},
  \bibinfo{author}{\bibfnamefont{A.~R.} \bibnamefont{Miles}},
  \bibinfo{author}{\bibfnamefont{S.~R.} \bibnamefont{Nagel}},
  \bibinfo{author}{\bibfnamefont{K.~S.} \bibnamefont{Raman}},
  \bibnamefont{et~al.}, {``}\bibinfo{title}{Rayleigh--Taylor instabilities in
  high-energy density settings on the National Ignition Facility},{''}
  \bibinfo{journal}{Proceedings of the National Academy of Sciences} p.
  \bibinfo{pages}{201717236} (\bibinfo{year}{2018}).

\bibitem[{\citenamefont{Campos and Morgan}(2019{\natexlab{a}})}]{Campos2019a}
\bibinfo{author}{\bibfnamefont{A.}~\bibnamefont{Campos}} \bibnamefont{and}
  \bibinfo{author}{\bibfnamefont{B.~E.} \bibnamefont{Morgan}},
  {``}\bibinfo{title}{Self-consistent feedback mechanism for the sudden viscous
  dissipation of finite-Mach-number compressing turbulence},{''}
  \bibinfo{journal}{Phys. Rev. E} \textbf{\bibinfo{volume}{99}},
  \bibinfo{pages}{013107} (\bibinfo{year}{2019}{\natexlab{a}}),
  \urlprefix\url{https://link.aps.org/doi/10.1103/PhysRevE.99.013107}.

\bibitem[{\citenamefont{{Braginskii}}(1965)}]{Braginskii1965}
\bibinfo{author}{\bibfnamefont{S.~I.} \bibnamefont{{Braginskii}}},
  {``}\bibinfo{title}{{Transport Processes in a Plasma}},{''}
  \bibinfo{journal}{Reviews of Plasma Physics} \textbf{\bibinfo{volume}{1}},
  \bibinfo{pages}{205} (\bibinfo{year}{1965}).

\bibitem[{\citenamefont{Cl\'erouin et~al.}(2020)\citenamefont{Cl\'erouin,
  Arnault, Gr\'ea, Guisset, Vandenboomgaerde, White, Collins, Kress, and
  Ticknor}}]{Clerouin2020}
\bibinfo{author}{\bibfnamefont{J.}~\bibnamefont{Cl\'erouin}},
  \bibinfo{author}{\bibfnamefont{P.}~\bibnamefont{Arnault}},
  \bibinfo{author}{\bibfnamefont{B.-J.} \bibnamefont{Gr\'ea}},
  \bibinfo{author}{\bibfnamefont{S.}~\bibnamefont{Guisset}},
  \bibinfo{author}{\bibfnamefont{M.}~\bibnamefont{Vandenboomgaerde}},
  \bibinfo{author}{\bibfnamefont{A.~J.} \bibnamefont{White}},
  \bibinfo{author}{\bibfnamefont{L.~A.} \bibnamefont{Collins}},
  \bibinfo{author}{\bibfnamefont{J.~D.} \bibnamefont{Kress}}, \bibnamefont{and}
  \bibinfo{author}{\bibfnamefont{C.}~\bibnamefont{Ticknor}},
  {``}\bibinfo{title}{Static and dynamic properties of multi-ionic plasma
  mixtures},{''} \bibinfo{journal}{Phys. Rev. E}
  \textbf{\bibinfo{volume}{101}}, \bibinfo{pages}{033207}
  (\bibinfo{year}{2020}),
  \urlprefix\url{https://link.aps.org/doi/10.1103/PhysRevE.101.033207}.

\bibitem[{\citenamefont{Davidovits and
  Fisch}(2016{\natexlab{a}})}]{Davidovits2016}
\bibinfo{author}{\bibfnamefont{S.}~\bibnamefont{Davidovits}} \bibnamefont{and}
  \bibinfo{author}{\bibfnamefont{N.~J.} \bibnamefont{Fisch}},
  {``}\bibinfo{title}{Sudden Viscous Dissipation of Compressing
  Turbulence},{''} \bibinfo{journal}{Phys. Rev. Lett.}
  \textbf{\bibinfo{volume}{116}}, \bibinfo{pages}{105004}
  (\bibinfo{year}{2016}{\natexlab{a}}),
  \urlprefix\url{https://link.aps.org/doi/10.1103/PhysRevLett.116.105004}.

\bibitem[{\citenamefont{Weber et~al.}(2014)\citenamefont{Weber, Clark, Cook,
  Busby, and Robey}}]{Weber2014}
\bibinfo{author}{\bibfnamefont{C.~R.} \bibnamefont{Weber}},
  \bibinfo{author}{\bibfnamefont{D.~S.} \bibnamefont{Clark}},
  \bibinfo{author}{\bibfnamefont{A.~W.} \bibnamefont{Cook}},
  \bibinfo{author}{\bibfnamefont{L.~E.} \bibnamefont{Busby}}, \bibnamefont{and}
  \bibinfo{author}{\bibfnamefont{H.~F.} \bibnamefont{Robey}},
  {``}\bibinfo{title}{Inhibition of turbulence in inertial-confinement-fusion
  hot spots by viscous dissipation},{''} \bibinfo{journal}{Phys. Rev. E}
  \textbf{\bibinfo{volume}{89}}, \bibinfo{pages}{053106}
  (\bibinfo{year}{2014}),
  \urlprefix\url{https://link.aps.org/doi/10.1103/PhysRevE.89.053106}.

\bibitem[{\citenamefont{Haines et~al.}(2016)\citenamefont{Haines, Grim, Fincke,
  Shah, Forrest, Silverstein, Marshall, Boswell, Fowler, Gore
  et~al.}}]{Haines2016}
\bibinfo{author}{\bibfnamefont{B.~M.} \bibnamefont{Haines}},
  \bibinfo{author}{\bibfnamefont{G.~P.} \bibnamefont{Grim}},
  \bibinfo{author}{\bibfnamefont{J.~R.} \bibnamefont{Fincke}},
  \bibinfo{author}{\bibfnamefont{R.~C.} \bibnamefont{Shah}},
  \bibinfo{author}{\bibfnamefont{C.~J.} \bibnamefont{Forrest}},
  \bibinfo{author}{\bibfnamefont{K.}~\bibnamefont{Silverstein}},
  \bibinfo{author}{\bibfnamefont{F.~J.} \bibnamefont{Marshall}},
  \bibinfo{author}{\bibfnamefont{M.}~\bibnamefont{Boswell}},
  \bibinfo{author}{\bibfnamefont{M.~M.} \bibnamefont{Fowler}},
  \bibinfo{author}{\bibfnamefont{R.~A.} \bibnamefont{Gore}},
  \bibnamefont{et~al.}, {``}\bibinfo{title}{Detailed high-resolution
  three-dimensional simulations of OMEGA separated reactants inertial
  confinement fusion experiments},{''} \bibinfo{journal}{Physics of Plasmas}
  \textbf{\bibinfo{volume}{23}}, \bibinfo{pages}{072709}
  (\bibinfo{year}{2016}), \eprint{https://doi.org/10.1063/1.4959117},
  \urlprefix\url{https://doi.org/10.1063/1.4959117}.

\bibitem[{\citenamefont{Zylstra et~al.}(2018)\citenamefont{Zylstra, Hoffman,
  Herrmann, Schmitt, Kim, Meaney, Leatherland, Gales, Forrest, Glebov
  et~al.}}]{Zylstra2018}
\bibinfo{author}{\bibfnamefont{A.~B.} \bibnamefont{Zylstra}},
  \bibinfo{author}{\bibfnamefont{N.~M.} \bibnamefont{Hoffman}},
  \bibinfo{author}{\bibfnamefont{H.~W.} \bibnamefont{Herrmann}},
  \bibinfo{author}{\bibfnamefont{M.~J.} \bibnamefont{Schmitt}},
  \bibinfo{author}{\bibfnamefont{Y.~H.} \bibnamefont{Kim}},
  \bibinfo{author}{\bibfnamefont{K.}~\bibnamefont{Meaney}},
  \bibinfo{author}{\bibfnamefont{A.}~\bibnamefont{Leatherland}},
  \bibinfo{author}{\bibfnamefont{S.}~\bibnamefont{Gales}},
  \bibinfo{author}{\bibfnamefont{C.}~\bibnamefont{Forrest}},
  \bibinfo{author}{\bibfnamefont{V.~Y.} \bibnamefont{Glebov}},
  \bibnamefont{et~al.}, {``}\bibinfo{title}{Diffusion-dominated mixing in
  moderate convergence implosions},{''} \bibinfo{journal}{Phys. Rev. E}
  \textbf{\bibinfo{volume}{97}}, \bibinfo{pages}{061201}
  (\bibinfo{year}{2018}),
  \urlprefix\url{https://link.aps.org/doi/10.1103/PhysRevE.97.061201}.

\bibitem[{\citenamefont{Viciconte et~al.}(2019)\citenamefont{Viciconte, Gr\'ea,
  Godeferd, Arnault, and Cl\'erouin}}]{Viciconte2019}
\bibinfo{author}{\bibfnamefont{G.}~\bibnamefont{Viciconte}},
  \bibinfo{author}{\bibfnamefont{B.-J.} \bibnamefont{Gr\'ea}},
  \bibinfo{author}{\bibfnamefont{F.~S.} \bibnamefont{Godeferd}},
  \bibinfo{author}{\bibfnamefont{P.}~\bibnamefont{Arnault}}, \bibnamefont{and}
  \bibinfo{author}{\bibfnamefont{J.}~\bibnamefont{Cl\'erouin}},
  {``}\bibinfo{title}{Sudden diffusion of turbulent mixing layers in weakly
  coupled plasmas under compression},{''} \bibinfo{journal}{Phys. Rev. E}
  \textbf{\bibinfo{volume}{100}}, \bibinfo{pages}{063205}
  (\bibinfo{year}{2019}),
  \urlprefix\url{https://link.aps.org/doi/10.1103/PhysRevE.100.063205}.

\bibitem[{\citenamefont{Campos and Morgan}(2019{\natexlab{b}})}]{Campos2019b}
\bibinfo{author}{\bibfnamefont{A.}~\bibnamefont{Campos}} \bibnamefont{and}
  \bibinfo{author}{\bibfnamefont{B.~E.} \bibnamefont{Morgan}},
  {``}\bibinfo{title}{Direct numerical simulation and Reynolds-averaged
  Navier-Stokes modeling of the sudden viscous dissipation for multicomponent
  turbulence},{''} \bibinfo{journal}{Phys. Rev. E}
  \textbf{\bibinfo{volume}{99}}, \bibinfo{pages}{063103}
  (\bibinfo{year}{2019}{\natexlab{b}}).

\bibitem[{\citenamefont{Mackay and Pino}(2020)}]{Mackay2020}
\bibinfo{author}{\bibfnamefont{K.~K.} \bibnamefont{Mackay}} \bibnamefont{and}
  \bibinfo{author}{\bibfnamefont{J.~E.} \bibnamefont{Pino}},
  {``}\bibinfo{title}{Modeling gas–shell mixing in ICF with separated
  reactants},{''} \bibinfo{journal}{Physics of Plasmas}
  \textbf{\bibinfo{volume}{27}}, \bibinfo{pages}{092704}
  (\bibinfo{year}{2020}), \eprint{https://doi.org/10.1063/5.0014856},
  \urlprefix\url{https://doi.org/10.1063/5.0014856}.

\bibitem[{\citenamefont{Haines et~al.}(2020)\citenamefont{Haines, Shah, Smidt,
  Albright, Cardenas, Douglas, Forrest, Glebov, Gunderson, Hamilton
  et~al.}}]{Haines2020}
\bibinfo{author}{\bibfnamefont{B.~M.} \bibnamefont{Haines}},
  \bibinfo{author}{\bibfnamefont{R.~C.} \bibnamefont{Shah}},
  \bibinfo{author}{\bibfnamefont{J.~M.} \bibnamefont{Smidt}},
  \bibinfo{author}{\bibfnamefont{B.~J.} \bibnamefont{Albright}},
  \bibinfo{author}{\bibfnamefont{T.}~\bibnamefont{Cardenas}},
  \bibinfo{author}{\bibfnamefont{M.~R.} \bibnamefont{Douglas}},
  \bibinfo{author}{\bibfnamefont{C.}~\bibnamefont{Forrest}},
  \bibinfo{author}{\bibfnamefont{V.~Y.} \bibnamefont{Glebov}},
  \bibinfo{author}{\bibfnamefont{M.~A.} \bibnamefont{Gunderson}},
  \bibinfo{author}{\bibfnamefont{C.}~\bibnamefont{Hamilton}},
  \bibnamefont{et~al.}, {``}\bibinfo{title}{The rate of development of atomic
  mixing and temperature equilibration in inertial confinement fusion
  implosions},{''} \bibinfo{journal}{Physics of Plasmas}
  \textbf{\bibinfo{volume}{27}}, \bibinfo{pages}{102701}
  (\bibinfo{year}{2020}), \eprint{https://doi.org/10.1063/5.0013456},
  \urlprefix\url{https://doi.org/10.1063/5.0013456}.

\bibitem[{\citenamefont{Lumley}(1992)}]{Lumley1992}
\bibinfo{author}{\bibfnamefont{J.~L.} \bibnamefont{Lumley}},
  {``}\bibinfo{title}{Some comments on turbulence},{''}
  \bibinfo{journal}{Physics of Fluids A: Fluid Dynamics}
  \textbf{\bibinfo{volume}{4}}, \bibinfo{pages}{203} (\bibinfo{year}{1992}),
  \eprint{https://doi.org/10.1063/1.858347},
  \urlprefix\url{https://doi.org/10.1063/1.858347}.

\bibitem[{\citenamefont{Taylor}(1935)}]{Taylor1935}
\bibinfo{author}{\bibfnamefont{G.~I.} \bibnamefont{Taylor}},
  {``}\bibinfo{title}{Statistical theory of turbulence},{''}
  \bibinfo{journal}{Proceedings of the Royal Society of London. Series A -
  Mathematical and Physical Sciences} \textbf{\bibinfo{volume}{151}},
  \bibinfo{pages}{421} (\bibinfo{year}{1935}),
  \eprint{https://royalsocietypublishing.org/doi/pdf/10.1098/rspa.1935.0158},
  \urlprefix\url{https://royalsocietypublishing.org/doi/abs/10.1098/rspa.1935.0158}.

\bibitem[{\citenamefont{Tennekes and Lumley}(1972)}]{Tennekes1972}
\bibinfo{author}{\bibfnamefont{H.}~\bibnamefont{Tennekes}} \bibnamefont{and}
  \bibinfo{author}{\bibfnamefont{J.~L.} \bibnamefont{Lumley}},
  \emph{\bibinfo{title}{A first course in turbulence}} (\bibinfo{publisher}{MIT
  press}, \bibinfo{year}{1972}).

\bibitem[{\citenamefont{Vassilicos}(2015)}]{Vassilicos2015}
\bibinfo{author}{\bibfnamefont{J.~C.} \bibnamefont{Vassilicos}},
  {``}\bibinfo{title}{Dissipation in Turbulent Flows},{''}
  \bibinfo{journal}{Annual Review of Fluid Mechanics}
  \textbf{\bibinfo{volume}{47}}, \bibinfo{pages}{95} (\bibinfo{year}{2015}),
  \eprint{https://doi.org/10.1146/annurev-fluid-010814-014637},
  \urlprefix\url{https://doi.org/10.1146/annurev-fluid-010814-014637}.

\bibitem[{\citenamefont{Davidovits and Fisch}(2017)}]{Davidovits2017}
\bibinfo{author}{\bibfnamefont{S.}~\bibnamefont{Davidovits}} \bibnamefont{and}
  \bibinfo{author}{\bibfnamefont{N.~J.} \bibnamefont{Fisch}},
  {``}\bibinfo{title}{Modeling turbulent energy behavior and sudden viscous
  dissipation in compressing plasma turbulence},{''} \bibinfo{journal}{Physics
  of Plasmas} \textbf{\bibinfo{volume}{24}}, \bibinfo{pages}{122311}
  (\bibinfo{year}{2017}), \eprint{https://doi.org/10.1063/1.5006946},
  \urlprefix\url{https://doi.org/10.1063/1.5006946}.

\bibitem[{\citenamefont{Lohse}(1994)}]{Loshe1994}
\bibinfo{author}{\bibfnamefont{D.}~\bibnamefont{Lohse}},
  {``}\bibinfo{title}{Crossover from High to Low Reynolds Number
  Turbulence},{''} \bibinfo{journal}{Phys. Rev. Lett.}
  \textbf{\bibinfo{volume}{73}}, \bibinfo{pages}{3223} (\bibinfo{year}{1994}),
  \urlprefix\url{https://link.aps.org/doi/10.1103/PhysRevLett.73.3223}.

\bibitem[{\citenamefont{Nishitani and Ishii}(1985)}]{Nishitani1985}
\bibinfo{author}{\bibfnamefont{T.}~\bibnamefont{Nishitani}} \bibnamefont{and}
  \bibinfo{author}{\bibfnamefont{K.}~\bibnamefont{Ishii}},
  {``}\bibinfo{title}{Similarity transformations of the {N}avier-{S}tokes
  equation},{''} \bibinfo{journal}{J. Phys. Soc. Japan}
  \textbf{\bibinfo{volume}{54}}, \bibinfo{pages}{5461} (\bibinfo{year}{1985}).

\bibitem[{\citenamefont{Cambon et~al.}(1992)\citenamefont{Cambon, Mao, and
  Jeandel}}]{Cambon1992}
\bibinfo{author}{\bibfnamefont{C.}~\bibnamefont{Cambon}},
  \bibinfo{author}{\bibfnamefont{Y.}~\bibnamefont{Mao}}, \bibnamefont{and}
  \bibinfo{author}{\bibfnamefont{D.}~\bibnamefont{Jeandel}},
  {``}\bibinfo{title}{On the application of time dependent scaling to the
  modelling of turbulence undergoing compression},{''} \bibinfo{journal}{Eur.
  J. Mech. B-Fluids.} \textbf{\bibinfo{volume}{11}}, \bibinfo{pages}{683–703}
  (\bibinfo{year}{1992}).

\bibitem[{\citenamefont{Cambon et~al.}(1993)\citenamefont{Cambon, Coleman, and
  Mansour}}]{Cambon1993}
\bibinfo{author}{\bibfnamefont{C.}~\bibnamefont{Cambon}},
  \bibinfo{author}{\bibfnamefont{G.~N.} \bibnamefont{Coleman}},
  \bibnamefont{and} \bibinfo{author}{\bibfnamefont{N.~N.}
  \bibnamefont{Mansour}}, {``}\bibinfo{title}{Rapid distortion analysis and
  direct simulation of compressible homogeneous turbulence at finite {M}ach
  number},{''} \bibinfo{journal}{Journal of Fluid Mechanics}
  \textbf{\bibinfo{volume}{257}}, \bibinfo{pages}{641–665}
  (\bibinfo{year}{1993}).

\bibitem[{\citenamefont{Orszag}(1970)}]{Orszag1970}
\bibinfo{author}{\bibfnamefont{S.~A.} \bibnamefont{Orszag}},
  {``}\bibinfo{title}{Analytical theories of turbulence},{''}
  \bibinfo{journal}{J. Fluid Mech.} \textbf{\bibinfo{volume}{41}},
  \bibinfo{pages}{363} (\bibinfo{year}{1970}).

\bibitem[{\citenamefont{Lesieur and Ossia}(2000)}]{Lesieur2000}
\bibinfo{author}{\bibfnamefont{M.}~\bibnamefont{Lesieur}} \bibnamefont{and}
  \bibinfo{author}{\bibfnamefont{S.}~\bibnamefont{Ossia}},
  {``}\bibinfo{title}{{3D} isotropic turbulence at very high {R}eynolds
  numbers: {EDQNM} study},{''} \bibinfo{journal}{J. Turb.}
  \textbf{\bibinfo{volume}{1}}, \bibinfo{pages}{1} (\bibinfo{year}{2000}),
  \eprint{http://www.tandfonline.com/doi/pdf/10.1088/1468-5248/1/1/007},
  \urlprefix\url{http://www.tandfonline.com/doi/abs/10.1088/1468-5248/1/1/007}.

\bibitem[{\citenamefont{Davidovits and
  Fisch}(2016{\natexlab{b}})}]{Davidovits2016b}
\bibinfo{author}{\bibfnamefont{S.}~\bibnamefont{Davidovits}} \bibnamefont{and}
  \bibinfo{author}{\bibfnamefont{N.~J.} \bibnamefont{Fisch}},
  {``}\bibinfo{title}{Compressing turbulence and sudden viscous dissipation
  with compression-dependent ionization state},{''} \bibinfo{journal}{Phys.
  Rev. E} \textbf{\bibinfo{volume}{94}}, \bibinfo{pages}{053206}
  (\bibinfo{year}{2016}{\natexlab{b}}).

\bibitem[{\citenamefont{Lesieur}(1990)}]{Lesieur1990}
\bibinfo{author}{\bibfnamefont{M.~R.} \bibnamefont{Lesieur}},
  \emph{\bibinfo{title}{Turbulence in fluids}} (\bibinfo{year}{1990}).

\bibitem[{\citenamefont{Pope}(2000)}]{Pope2000}
\bibinfo{author}{\bibfnamefont{S.~B.} \bibnamefont{Pope}},
  \emph{\bibinfo{title}{Turbulent Flows}} (\bibinfo{publisher}{Cambridge
  University Press}, \bibinfo{year}{2000}), ISBN \bibinfo{isbn}{9780521598866},
  \urlprefix\url{http://books.google.fr/books?id=HZsTw9SMx-0C}.

\bibitem[{\citenamefont{Perot and BruynKops}(2006)}]{Perot2006}
\bibinfo{author}{\bibfnamefont{J.~B.} \bibnamefont{Perot}} \bibnamefont{and}
  \bibinfo{author}{\bibfnamefont{S.~M.~D.} \bibnamefont{BruynKops}},
  {``}\bibinfo{title}{Modeling turbulent dissipation at low and moderate
  Reynolds numbers},{''} \bibinfo{journal}{Journal of Turbulence}
  \textbf{\bibinfo{volume}{7}} (\bibinfo{year}{2006}).

\bibitem[{\citenamefont{Jones and Launder}(1972)}]{Jones1972}
\bibinfo{author}{\bibfnamefont{W.~P.} \bibnamefont{Jones}} \bibnamefont{and}
  \bibinfo{author}{\bibfnamefont{B.~E.} \bibnamefont{Launder}},
  {``}\bibinfo{title}{The prediction of laminarization with a two-equation
  model of turbulence},{''} \bibinfo{journal}{International Journal of Heat and
  Mass Transfer} \textbf{\bibinfo{volume}{15}}, \bibinfo{pages}{301}
  (\bibinfo{year}{1972}).

\bibitem[{\citenamefont{Durbin}(1991)}]{Durbin1991}
\bibinfo{author}{\bibfnamefont{P.}~\bibnamefont{Durbin}},
  {``}\bibinfo{title}{Near-wall turbulence closure modeling without damping
  functions},{''} \bibinfo{journal}{Theoretical and Computational Fluid
  Dynamics} \textbf{\bibinfo{volume}{3}}, \bibinfo{pages}{1}
  (\bibinfo{year}{1991}).

\bibitem[{\citenamefont{Hanjalic and Jakirlic}(1998)}]{Hanjalic1998}
\bibinfo{author}{\bibfnamefont{K.}~\bibnamefont{Hanjalic}} \bibnamefont{and}
  \bibinfo{author}{\bibfnamefont{S.}~\bibnamefont{Jakirlic}},
  {``}\bibinfo{title}{Contribution towards the second-moment closure modeling
  of separating turbulent flows},{''} \bibinfo{journal}{Computers and Fluids}
  \textbf{\bibinfo{volume}{37}}, \bibinfo{pages}{147} (\bibinfo{year}{1998}).

\bibitem[{\citenamefont{Duraisamy et~al.}(2019)\citenamefont{Duraisamy,
  Iaccarino, and Xiao}}]{Duraisamy2019}
\bibinfo{author}{\bibfnamefont{K.}~\bibnamefont{Duraisamy}},
  \bibinfo{author}{\bibfnamefont{G.}~\bibnamefont{Iaccarino}},
  \bibnamefont{and} \bibinfo{author}{\bibfnamefont{H.}~\bibnamefont{Xiao}},
  {``}\bibinfo{title}{Turbulence modeling in the age of data},{''}
  \bibinfo{journal}{Annu. Rev. Fluid Mech.} \textbf{\bibinfo{volume}{51}}
  (\bibinfo{year}{2019}).

\bibitem[{\citenamefont{Brenner et~al.}(2019)\citenamefont{Brenner, Eldredge,
  and Freund}}]{Brenner2019}
\bibinfo{author}{\bibfnamefont{M.~P.} \bibnamefont{Brenner}},
  \bibinfo{author}{\bibfnamefont{J.~D.} \bibnamefont{Eldredge}},
  \bibnamefont{and} \bibinfo{author}{\bibfnamefont{J.~B.}
  \bibnamefont{Freund}}, {``}\bibinfo{title}{Perspective on machine learning
  for advancing fluid mechanics},{''} \bibinfo{journal}{Physical Review Fluids}
  \textbf{\bibinfo{volume}{4}} (\bibinfo{year}{2019}).

\bibitem[{\citenamefont{Brunton et~al.}(2020)\citenamefont{Brunton, Noack, and
  Koumoutsakos}}]{brunton_machine_2020}
\bibinfo{author}{\bibfnamefont{S.~L.} \bibnamefont{Brunton}},
  \bibinfo{author}{\bibfnamefont{B.~R.} \bibnamefont{Noack}}, \bibnamefont{and}
  \bibinfo{author}{\bibfnamefont{P.}~\bibnamefont{Koumoutsakos}},
  {``}\bibinfo{title}{Machine {Learning} for {Fluid} {Mechanics}},{''}
  \bibinfo{journal}{Annual Review of Fluid Mechanics}
  \textbf{\bibinfo{volume}{52}}, \bibinfo{pages}{477} (\bibinfo{year}{2020}),
  ISSN \bibinfo{issn}{0066-4189, 1545-4479},
  \urlprefix\url{https://www.annualreviews.org/doi/10.1146/annurev-fluid-010719-060214}.

\bibitem[{\citenamefont{Ling et~al.}(2016)\citenamefont{Ling, Jones, and
  Templeton}}]{Ling2016}
\bibinfo{author}{\bibfnamefont{J.}~\bibnamefont{Ling}},
  \bibinfo{author}{\bibfnamefont{R.}~\bibnamefont{Jones}}, \bibnamefont{and}
  \bibinfo{author}{\bibfnamefont{J.}~\bibnamefont{Templeton}},
  {``}\bibinfo{title}{Machine learning strategies for systems with invariance
  properties},{''} \bibinfo{journal}{J. Comp. Phys.}
  \textbf{\bibinfo{volume}{318}} (\bibinfo{year}{2016}).

\bibitem[{\citenamefont{Wu et~al.}(2018)\citenamefont{Wu, Xiao, and
  Paterson}}]{wu_physics-informed_2018}
\bibinfo{author}{\bibfnamefont{J.-L.} \bibnamefont{Wu}},
  \bibinfo{author}{\bibfnamefont{H.}~\bibnamefont{Xiao}}, \bibnamefont{and}
  \bibinfo{author}{\bibfnamefont{E.}~\bibnamefont{Paterson}},
  {``}\bibinfo{title}{Physics-informed machine learning approach for augmenting
  turbulence models: {A} comprehensive framework},{''}
  \bibinfo{journal}{Physical Review Fluids} \textbf{\bibinfo{volume}{3}},
  \bibinfo{pages}{074602} (\bibinfo{year}{2018}), \bibinfo{note}{publisher:
  American Physical Society},
  \urlprefix\url{https://link.aps.org/doi/10.1103/PhysRevFluids.3.074602}.

\bibitem[{\citenamefont{Karniadakis et~al.}(2021)\citenamefont{Karniadakis,
  Kevrekidis, Lu, Perdikaris, Wang, and
  Yang}}]{karniadakis_physics-informed_2021}
\bibinfo{author}{\bibfnamefont{G.~E.} \bibnamefont{Karniadakis}},
  \bibinfo{author}{\bibfnamefont{I.~G.} \bibnamefont{Kevrekidis}},
  \bibinfo{author}{\bibfnamefont{L.}~\bibnamefont{Lu}},
  \bibinfo{author}{\bibfnamefont{P.}~\bibnamefont{Perdikaris}},
  \bibinfo{author}{\bibfnamefont{S.}~\bibnamefont{Wang}}, \bibnamefont{and}
  \bibinfo{author}{\bibfnamefont{L.}~\bibnamefont{Yang}},
  {``}\bibinfo{title}{Physics-informed machine learning},{''}
  \bibinfo{journal}{Nature Reviews Physics} \textbf{\bibinfo{volume}{3}},
  \bibinfo{pages}{422} (\bibinfo{year}{2021}), ISSN \bibinfo{issn}{2522-5820},
  \urlprefix\url{https://www.nature.com/articles/s42254-021-00314-5}.

\bibitem[{\citenamefont{Brunton et~al.}(2016)\citenamefont{Brunton, Proctor,
  and Kutz}}]{Brunton2016}
\bibinfo{author}{\bibfnamefont{S.~L.} \bibnamefont{Brunton}},
  \bibinfo{author}{\bibfnamefont{J.~L.} \bibnamefont{Proctor}},
  \bibnamefont{and} \bibinfo{author}{\bibfnamefont{J.~N.} \bibnamefont{Kutz}},
  {``}\bibinfo{title}{Discovering governing equations from data by sparse
  identification of non linear dynamical systems},{''} \bibinfo{journal}{Proc.
  Nat. Acad. Sci. USA} \textbf{\bibinfo{volume}{113}}, \bibinfo{pages}{3932}
  (\bibinfo{year}{2016}).

\bibitem[{\citenamefont{Beetham and Capecelatro}(2020)}]{Beetham2020}
\bibinfo{author}{\bibfnamefont{S.}~\bibnamefont{Beetham}} \bibnamefont{and}
  \bibinfo{author}{\bibfnamefont{J.}~\bibnamefont{Capecelatro}},
  {``}\bibinfo{title}{Formulating turbulence closures using sparse regression
  with embedded form invariance},{''} \bibinfo{journal}{Physical Review Fluids}
  \textbf{\bibinfo{volume}{5}}, \bibinfo{pages}{084611} (\bibinfo{year}{2020}).

\end{thebibliography}
\end{document}